%% file: main.tex
\documentclass[traditabstract]{aa}
\usepackage     [utf8]                 {inputenc}  
\usepackage{verbatim} 
\usepackage{amsmath}
\usepackage{amssymb}
\usepackage{graphicx}
\usepackage{flafter} 
\usepackage{booktabs}
\usepackage[authoryear]{natbib}
\usepackage[colorlinks=true,linkcolor=blue,urlcolor=blue,citecolor=blue]{hyperref}
\usepackage{xspace}
\usepackage{siunitx}
\usepackage{lscape}
\usepackage{longtable}
\usepackage{threeparttable}

\makeatletter

\usepackage{txfonts}\usepackage{twoopt}
\bibpunct{(}{)}{;}{a}{}{,}
\usepackage{enumitem}

\usepackage[rightcaption]{sidecap}
\usepackage{graphicx} 
\graphicspath{ {images/} }

\newcommand{\msun}{{\rm M}_{\odot}}

\newcommand{\eg}{e.g.\@\xspace}
\newcommand{\cf}{c.f.\@\xspace}
\newcommand{\ie}{i.e.\@\xspace}
\newcommand{\hl}[1]{#1}

\titlerunning{Convective-core overshooting and the final fate of massive stars}
\authorrunning{D.~Temaj}
\makeatother
\begin{document}
\title{Convective-core overshooting and the final fate of massive stars}
\author{D.~Temaj\inst{\ref{HITS}}\thanks{duresa.temaj@gmail.com}, F.R.N.~Schneider\inst{\ref{HITS},\ref{ZAH}}, E.~Laplace\inst{\ref{HITS}}, D. Wei\inst{\ref{HITS}}, Ph.~Podsiadlowski\inst{\ref{Oxford},\ref{HITS}}}
\institute{
Heidelberger Institut f\"{u}r Theoretische Studien, Schloss-Wolfsbrunnenweg 35, 69118 Heidelberg, Germany\label{HITS}
\and Zentrum f\"{u}r Astronomie der Universit\"{a}t Heidelberg, Astronomisches Rechen-Institut, M\"{o}nchhofstr. 12-14, 69120 Heidelberg, Germany\label{ZAH}
\and University of Oxford, St Edmund Hall, Oxford, OX1 4AR, United Kingdom
 \label{Oxford}
}
\date{July 2023}
\abstract{
Massive stars can explode in powerful supernovae \hl{(SNe)} forming neutron stars but they may also collapse directly into black holes. Understanding and predicting their final fate is increasingly important, \eg, in the context of gravitational-wave astronomy. The interior mixing of stars in general and convective boundary mixing in particular remain some of the largest uncertainties in their evolution. Here, we investigate the influence of convective boundary mixing on the pre-supernova structure and explosion properties of massive stars. Using the 1D stellar evolution code \textsc{Mesa}, we model single, non-rotating stars of solar metallicity with initial masses of $5\,\text{--}\,70\,\mathrm{M_\odot}$ and convective core step-overshooting of $0.05\text{--}0.50$ pressure scale heights. Stars are evolved until the onset of iron core collapse, and the pre-SN models are exploded using a parametric, semi-analytic SN code. We use the compactness parameter to describe the interior structure of stars at core collapse and find a pronounced peak in compactness at carbon-oxygen core masses of $M_\mathrm{CO}\,{\approx}\,7\,\mathrm{M_\odot}$ and generally high compactness at $M_\mathrm{CO}\,{\gtrsim}\,14\,\mathrm{M_\odot}$. Larger convective core overshooting shifts the location of the compactness peak by $1\text{--}2\,\msun$ to higher $M_\mathrm{CO}$. These core masses correspond to initial masses of $24\,\msun$ ($19\,\msun$) and ${\gtrsim}\,40\,\msun$ (${\gtrsim}\,\hl{30}\,\msun$), respectively, in models with the lowest (highest) convective core overshooting parameter. In both high-compactness regimes, stars are found to collapse into black holes. As the luminosity of the pre-supernova progenitor is determined by $M_\mathrm{CO}$, we predict black-hole formation for progenitors with luminosities $5.35\,{\leq}\,\log(L/\mathrm{L_\odot})\,{\leq}\,5.50$ and $\log(L/\mathrm{L_\odot})\,{\geq}\,5.80$. The luminosity range of black-hole formation from stars in the compactness peak agrees well with the observed luminosity of the red supergiant star N6946\,BH1 that disappeared without a bright supernova and likely collapsed into a black hole. While some of our models in the luminosity range $\log(L/\mathrm{L_\odot})\,{=}\,5.1\,\text{--}\,5.5$ indeed collapse to form black holes, this does not fully explain the lack of observed SN~IIP progenitors at these luminosities, \ie the missing red-supergiant problem. The amount of convective boundary mixing also affects the wind mass loss of stars such that the lowest black hole mass are $15\,\msun$ and $10\,\msun$ in our models with \hl{the} lowest and highest convective core overshooting parameter, respectively. The compactness parameter, central specific entropy\hl{,} and iron core mass describe a qualitatively similar landscape as a function of $M_\mathrm{CO}$, and we find that entropy is a particularly good predictor of the neutron-star masses in our models. We find no correlation between the explosion energy, kick velocity, and nickel mass production with the convective core overshooting value, but a tight relation with the compactness parameter. We further show how convective core overshooting affects the pre-supernova locations of stars in the Hertzsprung--Russell diagram and the plateau luminosity and duration of SN~IIP lightcurves.
}

\maketitle
%
%
\section{Introduction}\label{sect: intro}

The feedback of massive stars via their strong winds, enormous radiation and powerful supernova \hl{(SN)} explosions \hl{play} an important role, \eg, in the evolution of galaxies and the chemical enrichment of the Universe \citep[\eg][]{burbidge_synthesis_1957,ceverino_role_2009, langer_presupernova_2012, nomoto_nucleosynthesis_2013, smith_mass_2014}. Which massive stars explode in SNe and which collapse into black holes \hl{(BHs)} without bright SNe is still an open question. Many studies have addressed this question by evolving stellar models to their pre-SN stage and exploring their explodability with SN simulations and parameterized SN codes \citep[\eg ][]{timmes_neutron_1996, woosley_evolution_2002, heger_how_2003, zhang_fallback_2008, oconnor_black_2011, ugliano_progenitor-explosion_2012, fryer_compact_2012, pejcha_landscape_2015, muller_simple_2016, ertl_two-parameter_2016, sukhbold_core-collapse_2016, sukhbold_high-resolution_2018, woosley_birth_2020, schneider_pre-supernova_2021, laplace_different_2021, patton_comparing_2022, aguilera-dena_stripped-envelope_2023}.
However, there are many uncertainties in massive star evolution that could affect the results of such studies \citep[see \eg][]{langer_presupernova_2012, maeder_rotating_2012}. 

One of the most considerable uncertainties in stellar evolution is convective boundary mixing, and it is required to, \eg, explain the width of the main sequence in the Hertzsprung–-Russell diagram (HRD), eclipsing binary stars and the core masses of stars as measured via asteroseismology \citep[\eg][]{chiosi_new_1992, schroder_critical_1997, langer_presupernova_2012, neiner_seismic_2012, stancliffe_confronting_2015, salaris_chemical_2017, claret_dependence_2019, pedersen_internal_2021, jermyn_convective_2022}. Also, 3D simulations show an additional mixing above the Schwarzschild boundary into the convectively stable layers, broadly in agreement with observations \citep[]{meakin_turbulent_2007, horst_multidimensional_2021, anders_stellar_2022, andrassy_dynamics_2022}.

Convective boundary mixing plays an important role in the evolution of stars \citep[\eg][]{schroder_critical_1997, langer_presupernova_2012, salaris_chemical_2017, higgins_massive_2019}. Because of the additional mixing into the convective burning core, there is more fuel to burn and the main sequence lifetime is prolonged. The mixing increases the mean molecular weight in the star resulting in a higher luminosity and cooler effective temperature. More convective boundary mixing causes the core to grow in mass, hence the star behaves as if it had a higher initial mass. While the effect of convective core overshooting has been extensively studied on the main sequence \citep[\eg ,][]{brott_rotating_2011,schootemeijer_constraining_2019}, the effects on the later evolutionary stages, and in particular on the final fate of massive stars and the properties of possible SNe have not been explored systematically.

Convective boundary mixing is essential in linking observations of SN progenitors to theoretical models. To date, tens of SN progenitors have been discovered, and there is a tension between the number of observed and expected SN~IIP progenitors, known as the missing red supergiant (RSG) problem \citep[][]{smartt_progenitors_2009}. There are no observed SN~IIP progenitors with luminosities higher than $\log(L/\mathrm{L_\odot})\,{\approx}\,5.1$ \citep{smartt_progenitors_2009,smartt_observational_2015} while RSGs are observed up to luminosities of $\log(L/\mathrm{L_\odot})\,{\approx}\,5.5$ \citep{davies_initial_2018}. The luminosity threshold of $\log(L/\mathrm{L_\odot})\,{\gtrsim}\,5.1$ at which no SN~IIP progenitors are found is commonly translated to initial masses of ${\gtrsim}\,18\,\mathrm{M_\odot}$ \citep[\eg][]{smartt_progenitors_2009, smartt_observational_2015}, but this conversion depends crucially on the applied stellar models and their assumptions, in particular regarding convective boundary mixing \citep[\eg ,][]{farrell_uncertain_2020}. One of the proposed solutions to the missing RSG problem is that RSGs more massive than ${\approx}\,18\,\mathrm{M_\odot}$ collapse into BHs without a bright SN and are thus not found in SN searches \citep{smartt_progenitors_2009, smartt_observational_2015}.

Here, we systematically explore how the extent of convective boundary mixing affects the pre-SN structure, the explodability of stars, the resulting SN explosion properties and the masses of the forming neutron stars (NSs) and BHs. We further investigate whether our models may help resolve the missing RSG problem. 
 
This paper is structured as follows. We describe our stellar evolution computations and the applied SN code in Sect.~\ref{sect: methods}. The pre-SN stellar structures of our models with different convective core overshooting values are shown in Sect.~\ref{sect: pre-sn evolution} and their predicted final fates in Sect.~\ref{sect: final fate}. We present the compact remnant masses left behind in Sect.~\ref{sect: Compact-remnants masses} and the explosion properties of our models in Sect.~\ref{sect:  Supernova explosion properties}. In Sect.~\ref{sect: HRD and missing RSG problem}, we show the models at the pre-SN stage in HRD and compare them to observed SN progenitors. We discuss our findings in Sect.~\ref{Discussion} and conclude in Sect.~\ref{Conclusion}.

\section{Methods}\label{sect: methods}

We evolve stellar models using the 1D stellar evolution code \textsc{Mesa} \hl{\citep[Modules for Experiments in Stellar Astrophysics,][]{paxton_modules_2011,paxton_modules_2013,paxton_modules_2015}.} Our grid contains a set of 250 models with initial masses in the range of $5\,{-}\,70\, \mathrm{\mathrm{M_\odot}}$. The evolution is carried out from the zero-age main sequence until the onset of the iron-core collapse (defined as when the infall velocity of the iron core reaches $950\, \mathrm{km\,s^{-1}}$). Due to numerical difficulties which we encountered, some of the lower-mass models (lower than ${\approx}\,9\,\mathrm{M_\odot}$) in our grid could not be evolved this far in \textsc{Mesa}. Instead, we use the \cite{tauris_ultra-stripped_2015} criterion to predict their final fate, \ie models with a final core-mass\footnote{Following \citet{tauris_ultra-stripped_2015}, final core mass is defined as where the helium mass fraction drops below $10\%$, which corresponds to the CO core mass in our models.} higher than $1.43\, \mathrm{M_\odot}$ produce iron core-collapse supernovae (Fe CCSN\hl{e}), those with a final core-mass between $1.37\,{-}\,1.43\, \mathrm{M_\odot}$ give rise to electron capture supernova\hl{e} (ECSN\hl{e}) and models with a final core-mass $ {<}\, 1.37\, \mathrm{M_\odot}$ form white dwarfs (WD).

Our physics assumptions for \textsc{Mesa} models are given in Sect.~\ref{sect: stellar evolution simulations with Mesa} and we introduce our SN code in Sect.~\ref{sect:  semi-analytic SN code}. All \textsc{Mesa} details and the data from the analysis with the SN code are given in Appendix \ref{appendix}.

\subsection{Stellar evolution simulations with \textsc{Mesa}}\label{sect: stellar evolution simulations with Mesa}
For the evolution of stellar models, we use \textsc{Mesa} version 10398. We evolve single, massive, non-rotating stars at solar metallicity \citep[$Z\,{=}\,0.0142$,][]{asplund_chemical_2009} with a helium mass fraction $Y{=}0.2703$. We follow the same setup as in \cite{schneider_pre-supernova_2021} but change the convective core overshooting parameter $\alpha_\mathrm{ov}\,{=}\,d_\mathrm{ov}/H_\mathrm{P}$ with $d_\mathrm{ov}$ being the distance to which the convective boundary mixing extends above the convective core, $H_\mathrm{P}$ the pressure-scale height and $\alpha_\mathrm{ov}$ the scaling parameter. To simulate convective boundary mixing, we use the step convective core overshooting scheme in \textsc{Mesa} with $\alpha_\mathrm{ov}$ values of $0.05, 0.10, 0.15, 0.20, 0.25, 0.30, 0.35,0.40, 0.45, 0.50$. We assume convective boundary mixing above convective core-hydrogen and core-helium burning. Convection is treated using the mixing length theory following \cite{henyey_studies_1965} with mixing length parameter $\alpha_\mathrm{MLT}\,{=}\,1.8$. For stability against convection, we employ the Ledoux criterion. To avoid numerical difficulties for stars approaching the Eddington limit, we enable MLT++. Semi-convection is applied with an efficiency parameter of $\alpha_\mathrm{sc}\,{=}\,0.1$ and a thermohaline mixing coefficient of 666 is adopted until core-oxygen exhaustion.

For stellar winds, we apply the custom \cite{schneider_pre-supernova_2021} wind-mass-loss scheme, that is similar to the `Dutch' wind mass-loss scheme but with modified metallicity scalings. For cool stars with effective temperature of $\log\,(T_\mathrm{eff}/\mathrm{K})\,{<}\,4$, we apply the \cite{nieuwenhuijzen_parametrization_1990} wind mass-loss rates. For hot stars with $\log\,(T_\mathrm{eff}/\mathrm{K})\,{>}\,4$ we apply the \cite{vink_new_2000,vink_mass-loss_2001} wind mass-loss rates. If the hydrogen mass fraction at the surface drops below $0.4$, we apply the Wolf-Rayet (WR) wind mass-loss rates of \cite{nugis_mass-loss_2000}. We enable a nuclear burning network of 21 base isotopes plus $\mathrm{^{56} Co}$ and $ \mathrm{^{60} Cr}$ (\textsc{approx21\_cr60\_plus\_co56.net}) in \textsc{Mesa} which covers all main burning stages. The reaction rates of $\mathrm{ ^{12} C (\alpha, \gamma) ^{16}O}$ are those of \cite{xu_nacre_2013}. The spatial and temporal evolution of these models is set with the \textsc{Mesa} controls \texttt{varcontrol\_target ${=}\,10^{-5}$} and \texttt{mesh\_delta\_coeff ${=}\,0.6$}, and we allow for the maximum cell size of \texttt{max\_dq ${=}\,10^{-3}$}.

We compute the gravitational binding energy $E_\mathrm{bind}$ of the matter above the iron core\footnote{Iron core includes all species with a mass number more than 46. The definition of iron core boundary by \textsc{Mesa} is when the $^{28}\mathrm{Si}$ mass-fraction is below 0.01 and the iron mass fraction is more than 0.1.} for our models at the onset of iron core collapse. 
The gravitational binding energy is given as 
\begin{equation}
    E_\mathrm{bind}\,{=}\,{-}\, \int_{M_\mathrm{Fe}}^{M_\mathrm{surf}}\,\frac{G\,m}{r} \mathrm{d}m, 
\end{equation}
the integration is from the iron core mass coordinate $M_\mathrm{Fe}$ to the surface of the star $M_\mathrm{surf}$, m is the mass coordinate and r the radius at that given mass coordinate.

Models that are left with a hydrogen envelope mass of more than $1\,\mathrm{M_\odot}$ at the onset of core-collapse are assumed to produce type IIP SNe, and the ones with a hydrogen envelope mass of $0.01\,{-}1\, \mathrm{M_\odot}$ to produce SN~IIb \citep{sravan_progenitors_2019}. Progenitor stars with a helium envelope mass of $0.06\,{-}\,0.14\, \mathrm{M_\odot}$ are assumed to give rise to SN~Ib/c \citep{hachinger_how_2012}.

\subsection{Semi-analytic SN code}\label{sect:  semi-analytic SN code}
We analyze the pre-SN stellar structure of our models that reach iron core collapse (Fe CC) using the semi-analytic parametric supernova code of \cite{muller_simple_2016}. The explosion energy $E_\mathrm{expl}$ is a combination of both the nuclear-burning energy and the neutrino-heating energy. We allow for a maximum NS mass of $2\, \mathrm{M_\odot}$, and for models with a remnant mass of more than $2\, \mathrm{M_\odot}$, a direct collapse to a black hole is assumed \citep{muller_simple_2016}. All material above a mass coordinate where the neutrino heating launches an explosion is ejected in a supernova explosion \citep{muller_simple_2016}. If after shock revival the explosion energy becomes negative at some mass coordinate, fallback material emerges, thereby reducing the ejected mass \citep{muller_simple_2016}. As in \cite{schneider_pre-supernova_2021}, we assume a fallback fraction of 50\% of the ejected material $M_\mathrm{ej}$ in such cases. If the accretion of the ejected material onto the proto-neutron star does not stop before it reaches a maximum NS mass of $2\, \mathrm{M_\odot}$ a black hole by fallback is formed. For models that launch a successful explosion, a constant asymmetry parameter is assumed, and the kick velocity of the NS is calculated as in \cite{schneider_pre-supernova_2021}.

Following the scalings of \cite{popov_analytical_1993}, we estimate the plateau luminosity $L_\mathrm{SN}$ and the duration of the plateau $t_\mathrm{SN}$ of type IIP SN light curve as 
\begin{equation}\label{eq: plateau L}
    L_\mathrm{SN} \,{\propto}\, E_\mathrm{expl}^{5/6} \,{\cdot}\, M_\mathrm{ej}^{-1/2} \,{\cdot}\, R^{2/3},
\end{equation}

\begin{equation}\label{eq: plateau t}
    t_\mathrm{SN} \,{\propto}\, E_\mathrm{expl}^{-1/6} \,{\cdot}\, M_\mathrm{ej}^{1/2} \,{\cdot}\, R^{1/6},
\end{equation}
with $R$ being the radius of the progenitor star at the onset of iron core collapse.


\section{Pre-SN stellar structure}\label{sect: pre-sn evolution}

\subsection{Interplay of convective core overshooting and wind mass loss in the evolution to core collapse}

\begin{SCfigure*}
            \centering
             \includegraphics{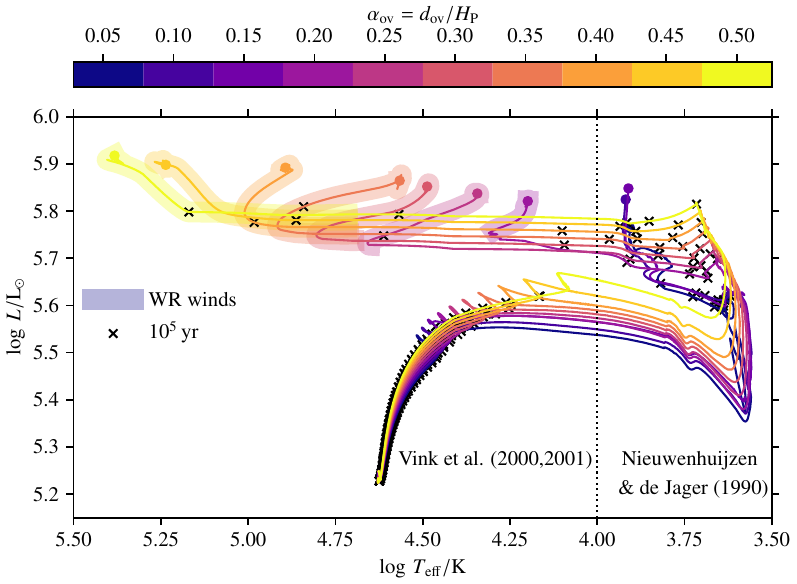}
            \caption{Evolution of initially $35\, \mathrm{M_\odot}$ models from the zero-age main sequence until the onset of iron core-collapse (filled circles). Colors show models of different convective core overshooting values. Black crosses indicate intervals of $10^5 \,\mathrm{yr}$. The black dotted line marks the transition from the hot to the cool wind mass loss regime at $\log (T_\mathrm{eff}/\mathrm{K})\,{=}\,4$.~Thick lines show where Wolf-Rayet wind mass-loss rates are applied.}
            \label{fig:  HRD ov}
\end{SCfigure*}

In Fig.~\ref{fig:  HRD ov}, we show the HRD of the evolution of a $35\, \mathrm{M_\odot}$ star with different convective core overshooting values shown with different colors. The evolution starts at the zero-age main sequence and ends at the onset of iron core-collapse (indicated by circles). Black crosses are separated by $10^5 \,\mathrm{yr}$ and show that models spend most of their time on the main sequence where the \cite{vink_new_2000,vink_mass-loss_2001} mass-loss rates are applied. After the main sequence, the envelope expands, and the star evolves to lower effective temperatures. When the effective temperature of the star reaches $\log\, (T_\mathrm{eff}/\mathrm{K})\,{<}\,4$, we switch to the stronger wind mass-loss rates of \citet[][see also Sect.~\ref{sect: methods}]{nieuwenhuijzen_parametrization_1990}. When the surface hydrogen mass fraction drops below $X_\mathrm{surf}\,{<}\,0.4$, we apply the WR wind mass-loss rates of \cite{nugis_mass-loss_2000} indicated by thick lines.

\hl{At the end of the main sequence evolution, the $35\,\mathrm{M_\odot}$ model with the highest convective core overshooting value reaches $\log\, (L/L_\odot)\,{\approx}\,5.63$ and $\log\, (T_\mathrm{eff}/\mathrm{K})\,{\approx}\,4.1$. It has a higher luminosity and cooler effective temperature compared to the model with the lowest convective core overshooting value, which at the end of the main sequence reaches values of $\log\, (L/L_\odot)\,{\approx}\,5.53$ and $\log\, (T_\mathrm{eff}/\mathrm{K})\,{\approx}\,4.5$.}
That is because the model with the highest convective core overshooting has a more extended mixing region and more hydrogen is mixed into the burning core which leads to an increase in the mean molecular weight of the star. \hl{Consequently, the star reaches a higher luminosity, and builds a more massive He core, leading to a larger radius and lower effective temperature}.
The $35\, \mathrm{M_\odot}$ model with the lowest convective core overshooting value spends its post-main-sequence lifetime in the cool wind mass loss regime and ends its life as a red supergiant. The models with a higher convective core overshooting value evolve to higher luminosities because they form more massive cores. The model with the highest convective core overshooting value ends its life with an effective temperature 30 times higher than the model with the lowest convective core overshooting. This can be explained by considering the mass loss in stellar winds during their evolution. Models with higher convective core overshooting form more massive cores and are left with a smaller envelope, thus wind mass loss exposes the hot interiors of the stars and pushes them blueward in the HRD. In the $35\, \mathrm{M_\odot}$ model, the \cite{nieuwenhuijzen_parametrization_1990} mass-loss rates are typically 10 times stronger than those of \cite{vink_new_2000,vink_mass-loss_2001}, and 3 times stronger than the WR rates.

\begin{figure*}
            \centering
             \includegraphics{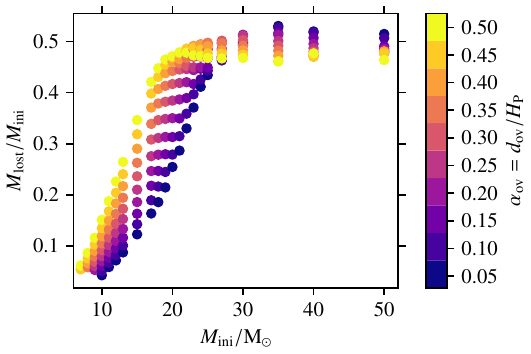}\hfill
             \includegraphics{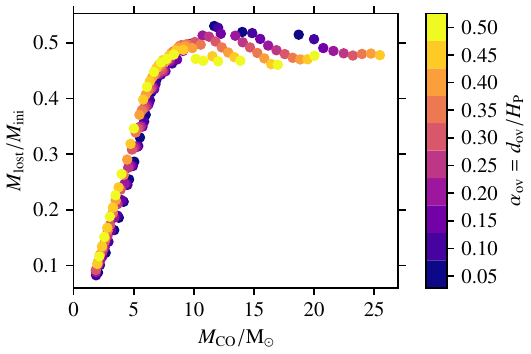}
            \caption{Total mass lost $M_\mathrm{lost}$ during the evolution relative to the initial mass $M_\mathrm{ini}$ of our entire set of models as a function of the initial mass (left) and CO core mass (right). Colors indicate different convective core overshooting values.}
            \label{fig: fractional mass loss}
\end{figure*}

In Fig.~\ref{fig: fractional mass loss}, we show the total mass lost by stars relative to their initial mass. We find that for initial masses up to $21\,\mathrm{M_\odot}$, models with the highest convective core overshooting lose the largest fraction of the mass, while this trend is reversed for initial masses greater than $30\, \mathrm{M_\odot}$. For a fixed convective core overshooting value we find a ``sweet spot'' where the fractional mass loss is the largest. This is found in initial mass models of $21\,\mathrm{M_\odot}$ and $35\,\mathrm{M_\odot}$ for the highest and lowest convective core overshooting value, respectively. The total mass loss of stars depends on the wind mass-loss rates and for how long stars are subject to which wind mass-loss regime.

Because the models with the highest convective core overshooting reach the highest luminosity, their mass-loss rates are the highest throughout all wind mass-loss regimes compared to the models with the lowest convective core overshooting value. During the post-main sequence evolution, models with initial masses ${\leq}\,21\,\mathrm{M_\odot}$ evolve quickly to cool effective temperatures and experience strong wind mass-loss rates until the end of their life.
For the $20\,\mathrm{M_\odot}$ model with the highest convective core overshooting, the RSG mass-loss rates reach up to $2\times 10^{-5} \,\mathrm{M_\odot\,yr^{-1}}$ while for the same initial mass but with the lowest convective core overshooting, RSGs mass loss rates are $6\times 10^{-6} \,\mathrm{M_\odot\,yr^{-1}}$.
The $20\, \mathrm{M_\odot}$ model with the highest convective core overshooting loses up to ${\approx}\,48 \,\%$ of its initial mass, while this is only $25\,\%$ for the lowest convective core overshooting model.

Models with initial masses ${\geq}30\, \mathrm{M_\odot}$ and the highest convective core overshooting value lose a smaller fraction of their mass compared to those with the lowest convective core overshooting.
The $35\, \mathrm{M_\odot}$ model with the lowest convective core overshooting spends $0.51\,\mathrm{Myr}$ on the cool wind mass loss regime, being subject to average wind mass-loss rates of ${\approx}\,3.2\times 10^{-5}\,\mathrm{M_\odot\,yr^{-1}}$. In total it loses $16.1\,\mathrm{M_\odot}$ due to RSG winds. For the $35\, \mathrm{M_\odot}$ model with the highest convective core overshooting, mass-loss rates in the cool regime on average are ${\approx}\,2.7\times 10^{-5}\,\mathrm{M_\odot\,yr^{-1}}$. Because of its large core, this model has a smaller envelope mass and its hot interiors are exposed after being subject to cool wind mass-loss rates for only $0.32\,\mathrm{Myr}$. The star evolves to higher effective temperatures and becomes a WR star whose wind mass-loss rates are smaller than RSGs winds. The $35\, \mathrm{M_\odot}$ model with the highest convective core overshooting is subject to WR winds for $0.1\,\mathrm{Myr}$ with average mass-loss rates of ${\approx}\,2.7\times 10^{-6}\,\mathrm{M_\odot\,yr^{-1}}$. 
Therefore, the total mass lost by the $35\, \mathrm{M_\odot}$ model with the highest convective core overshooting is smaller compared to the one with the lowest convective core overshooting.

In the right panel of Fig.~\ref{fig: fractional mass loss} we show the total mass lost by stars relative to their initial mass as a function of the CO core mass. We find that up to $M_\mathrm{CO}\,{\approx}\,10\, \mathrm{M_\odot}$ models with the same $M_\mathrm{CO}$ lose the same fraction of the initial mass. For models with CO core masses ${\geq}\,10\, \mathrm{M_\odot}$ they reach a plateau-like phase where all models lose ${\sim}\,50\%$ of their initial mass. Models with $M_\mathrm{CO}\,{>}\,10\,\mathrm{M_\odot}$ and the highest convective core overshooting value, lose a smaller fraction of the initial mass because they rapidly evolve through the strongest mass-loss region and become subject to WR winds as described previously. Based on this relation of the fractional mass loss with the CO core masses for our chosen RSG and WR mass-loss prescription, for a given CO core mass we can predict rather well how much the star's initial mass is going to be lost in stellar winds.

\subsection{Core properties}\label{sect: core properties}
        
\begin{figure}
            \centering
             \includegraphics{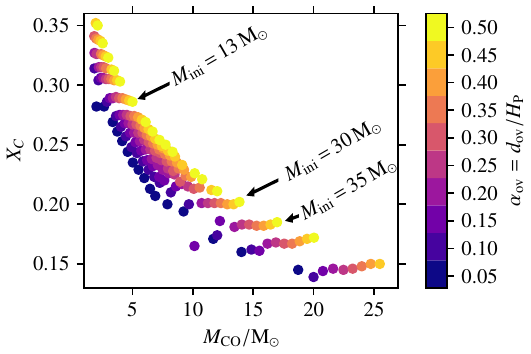}
            \caption{Core carbon mass fraction at the end of core-helium burning as a function of the CO core mass. Colors indicate different convective core overshooting values. We mark models with initial masses of $13\,\mathrm{M_\odot}$, $30\,\mathrm{M_\odot}$ and $35\,\mathrm{M_\odot}$.}
            \label{fig: carbon mass fraction}
\end{figure}

The ultimate fate of stars is determined by the conditions at the end of core helium burning \citep{chieffi_presupernova_2020,patton_towards_2020, schneider_pre-supernova_2021}, namely by the CO core mass and the carbon abundance in the core. During core-helium burning, two different nuclear burning processes take place: the triple-$\alpha$ process in which helium is converted to carbon, and $\alpha$-captures in which carbon is fused into oxygen via the $^{12}\mathrm{C}(\alpha,\gamma)^{16}\mathrm{O}$ nuclear reaction. Both reaction rates depend on the density, temperature, and amount of helium available in the core. The triple-$\alpha$ reaction rate scales with the third power of the $\alpha$ particle density, while $^{12}\mathrm{C}(\alpha,\gamma)^{16}\mathrm{O}$ scales linearly \citep{burbidge_synthesis_1957}.

In Fig.~\ref{fig: carbon mass fraction}, we show the \hl{core} carbon mass fraction of our models at the end of core helium burning as a function of the CO core mass, with colors showing different convective core overshooting values. High-mass models have a lower carbon abundance in the core because they have larger cores of lower ($\alpha$-particle) density; therefore $\alpha$-captures onto carbon are favored over triple-$\alpha$ and carbon is converted into oxygen more efficiently \citep{woosley_evolution_1995, brown_formation_1996, brown_formation_2001}.

In models with $M_\mathrm{ini}$ up to $13\,\mathrm{M_\odot}$ the carbon abundance slightly decreases when increasing the convective core overshooting value. That is because with increasing convective core overshooting value, models behave like more massive stars in the sense that they have more massive CO cores of low density (see Fig.~\ref{fig: carbon mass fraction}).  

For models with initial masses $13\,{-}\,30\,\mathrm{M_\odot}$, wind mass loss becomes more relevant and there is no clear trend with the convective core overshooting value. For the same initial mass model but different convective core overshooting, we find almost the same carbon abundance in the core. 

In the most massive stars ($\mathrm{M_{ini}} {>}\,30\, \mathrm{M_\odot}$), models with the highest convective core overshooting tend to have a higher carbon abundance in the CO core. The $35\,\mathrm{M_\odot}$ model with the lowest convective core overshooting forms a CO core mass of $11\,\mathrm{M_\odot}$ with a carbon abundance of $0.16$. The same initial mass model but with the highest convective core overshooting forms a CO core mass of $17\,\mathrm{M_\odot}$ with a carbon abundance of $0.18$. 
The high-mass models with the highest convective core overshooting lose most of the hydrogen-rich envelope in stellar winds, consequently, do not have a hydrogen-burning shell. As a response to the mass loss, the convective core during helium burning decreases in mass, leaving fewer $\alpha$-particles to be captured onto carbon and produce oxygen. Hence, the carbon abundance in the models with higher convective core overshooting is slightly higher. Similar results are reported by \cite{schneider_pre-supernova_2021} and \citet{laplace_different_2021} for binary-stripped stars. 

\begin{figure}
    \centering
     \includegraphics{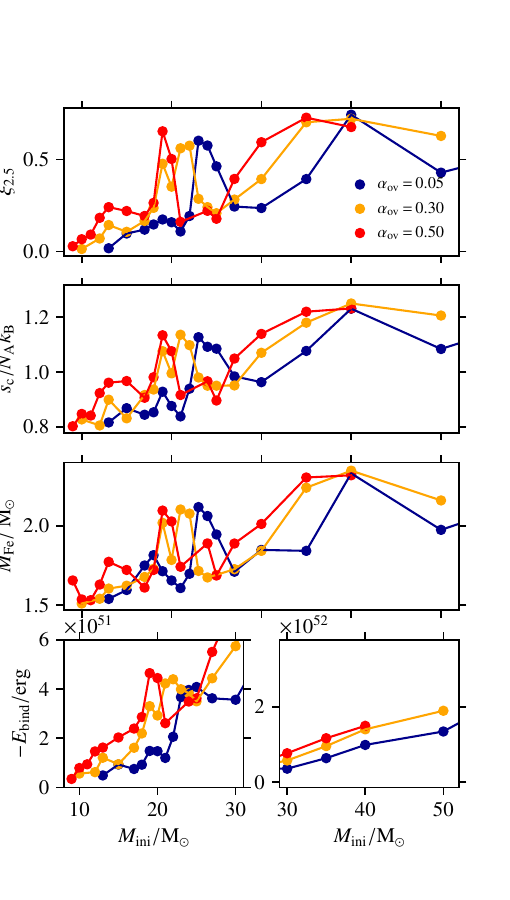}
    \caption{Final compactness parameter $\xi_{2.5}$, central specific entropy $s_\mathrm{c}$, iron core mass $M_\mathrm{Fe}$ and the binding energy above the iron core $E_\mathrm{bind}$ as a function of initial masses. Colors indicate models for different convective core overshooting parameters.}
    \label{fig: compactness vs Mini}
\end{figure}

To describe the core structure and infer the explodability of stars, the compactness parameter is often used. The compactness parameter $\xi_{M}$ is defined as 
\begin{equation}
    \xi_{M}=\frac{M/\mathrm{M_\odot}}{R(M)/\mathrm{1000\,km}},
\end{equation}
and is usually evaluated at a mass coordinate of $M\,{=}\,2.5\, \mathrm{M_\odot}$ \citep{oconnor_black_2011}. 
Models with a low compactness parameter \hl{($\boldsymbol{\xi_{M}}\boldsymbol{\leq} \boldsymbol{0.45}$)} tend to produce supernovae and leave behind NSs, and models with a high compactness parameter \hl{($\boldsymbol{\xi_{M}}\boldsymbol{\geq} \boldsymbol{0.45}$)} are more likely to implode and form BHs \citep{oconnor_black_2011}.
The compactness parameter follows a pattern as a function of initial mass (Fig.~\ref{fig: compactness vs Mini}) with the first peak at $19\,{-}\,24\, \mathrm{M_\odot}$ to which we refer as the first compactness peak and a second increase at initial masses of $30\,{-}\,40\, \mathrm{M_\odot}$ that we call the second compactness peak \citep{oconnor_black_2011, sukhbold_compactness_2014, sukhbold_high-resolution_2018, patton_towards_2020, chieffi_presupernova_2020}.
\hl{For the models with the highest convective core overshooting value $\alpha_\mathrm{ov}\,{=}\,0.50$ the first increase in compactness is at $19\,\mathrm{M_\odot}$ (with a compactness value of 0.65). After the first drop, the second increase in compactness is at the $30\,\mathrm{M_\odot}$ model (with a compactness value of 0.59).}
In the models at the first compactness peak core carbon burning proceeds radiatively, and for the models at the second peak of compactness core neon burning also turns radiative \citep{ schneider_pre-supernova_2021}.

In Fig.~\ref{fig: compactness vs Mini}, we show the final compactness parameter $\xi_\mathrm{2.5}$, central specific entropy $s_\mathrm{c}$, the iron core mass $M_\mathrm{Fe}$ and the binding energy above the iron core $E_\mathrm{bind}$ as a function of the initial mass for the models with three different convective core overshooting values ($\alpha_\mathrm{ov}\,{=}\,0.05, 0.30$ and $0.50$). The central entropy and the iron core masses follow the same trend as the compactness parameter as a function of the initial masses \citep{schneider_pre-supernova_2021, takahashi_monotonicity_2023}. We find that the gravitational binding energy also follows the same non-monotonic trend as compactness as a function of initial mass with a first peak at $19\,{-}\,24\, \mathrm{M_\odot}$ and after the first peak almost a linear increase in the binding energy.
By increasing the convective core overshooting parameter, the peaks are shifted to lower initial masses. 
For models with the lowest convective core overshooting value, the first and second peaks are at $M_\mathrm{ini}\,{\approx}\, 24\, \mathrm{M_\odot}$ and at $M_\mathrm{ini}\,{\approx}\, 40\, \mathrm{M_\odot}$. For the models with the highest convective core overshooting value, the peaks are shifted for $5\, \mathrm{M_\odot}$ and $10\, \mathrm{M_\odot}$ towards lower initial masses, and are at $M_\mathrm{ini}\,{\approx}\,19\,\mathrm{M_\odot}$ and $M_\mathrm{ini}\,{\approx}\,30\,\mathrm{M_\odot}$ for the first and second peak, respectively (see Fig.~\ref{fig: compactness vs Mini}). Lower-mass models with a high convective core overshooting value behave like the more massive stars with lower convective core overshooting and thus they have similar CO core masses.

\begin{figure}
    \centering
     \includegraphics{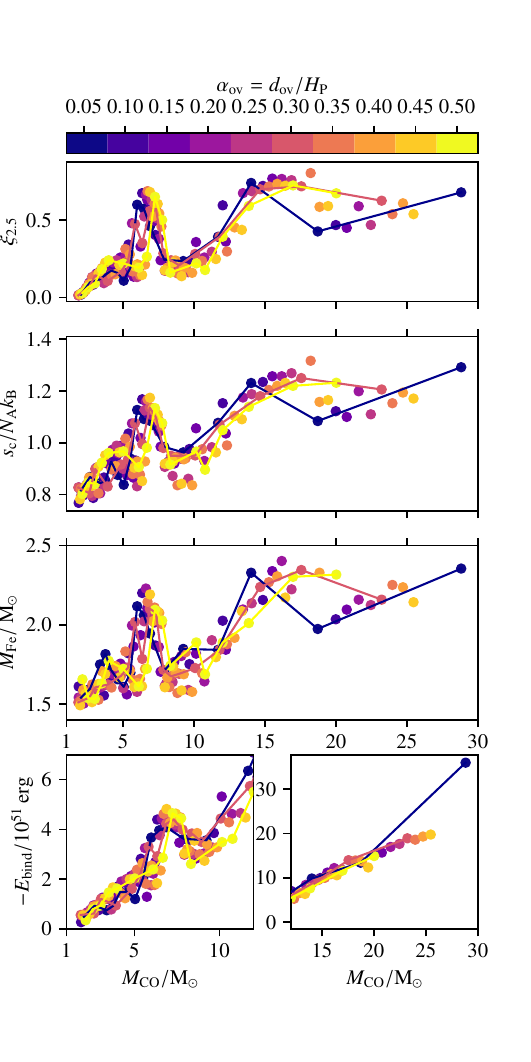}
    \caption{Similar to Fig.~\ref{fig: compactness vs Mini} but as a function of the CO core mass at the end of core He burning for all models. Colors show models of different convective core overshooting values. The blue, pink, and yellow lines connect models with the same convective core overshooting value for $\alpha_\mathrm{ov}\,{=}\,0.05, 0.30$ and $0.50$. }
    \label{fig: compactness vs Mco}
\end{figure}

The final compactness parameter, central specific entropy, the iron core mass, and the binding energy above the iron core are shown in Fig.~\ref{fig: compactness vs Mco} as a function of their CO core mass. Independently of the convective core overshooting value, we find the first compactness peak at the CO core masses of ${\sim}\,7\, \mathrm{M_\odot}$, and the second peak at ${\sim}\,14\, \mathrm{M_\odot}$, except for the binding energy after $M_\mathrm{CO}\,{\geq}\,10\,\mathrm{M_\odot}$ there is no second peak but almost a linear increase with $M_\mathrm{CO}$ (see Fig.~\ref{fig: compactness vs Mco}). However, for the highest convective core overshooting value, there is a slight shift of the peaks toward higher CO core masses, because initially lower mass models with the highest convective core overshooting value form a slightly larger CO core mass than the higher mass models with the lowest convective core overshooting value. Similar trend and binding energy values we find when we compute the binding energy of our models above the $M_4$ (the mass coordinate at a specific entropy of $s\,{=}\,4$) as our inner mass boundary.
Similarly to the initial masses, the central entropy, and the iron core masses follow the same trend as the compactness parameter with the CO core mass \citep{schneider_pre-supernova_2021} as well as the gravitational binding energy. All these quantities hold information about the internal structure of the star and are often used as a proxy to infer the explodability of the stars. Models at $M_\mathrm{CO}\,{\sim}\,7\, \mathrm{M_\odot}$ have a jump in the binding energy by a factor of 3 which shows that these models not only have high compactness, high central specific entropy, and high iron core but also a high binding energy which makes these stars more difficult to explode. Therefore these models are expected to collapse and form BHs.      

\section{Final Fate}\label{sect: final fate}

\begin{figure}
    \centering 
     \includegraphics{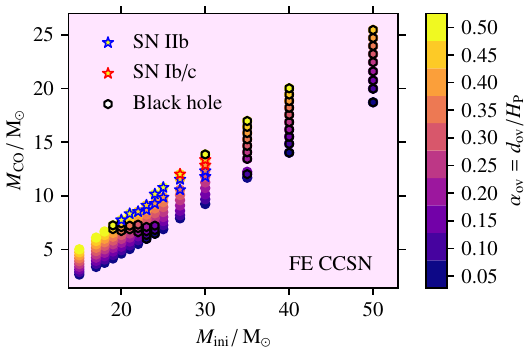}\hfill
     \includegraphics{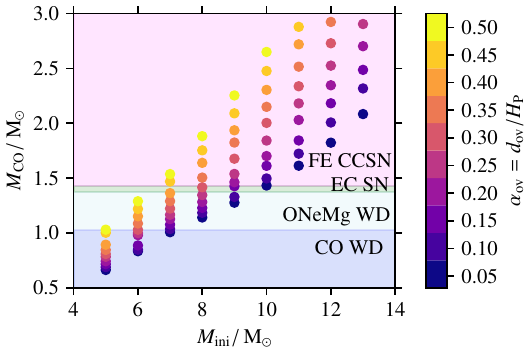}
    \caption{Carbon-oxygen core mass as a function of the initial masses for all our models in initial mass ranges of $5\,{-}\,14\, \mathrm{M_\odot}$ (bottom), and $14\,{-}\,50\, \mathrm{M_\odot}$ (top).~Colors show the convective core overshooting values.~The pink, green, sky-blue, and dark-blue regions show the regions of $M_\mathrm{CO}$ where iron core-collapse supernovae, electron capture supernovae, and ONeMg white dwarfs are expected following \cite{tauris_ultra-stripped_2015}. The blue-shaded region shows the CO core masses of models that could not reach carbon ignition thus CO white dwarfs are assumed to form. The blue and red-edged star symbols indicate the models that produce SN~IIb and SN~Ib/c.~\hl{Black-edged hexagonal shapes show models that form BHs}.}
    \label{fig: SN types}
\end{figure}

For the models that do not reach Fe CC, we predict their final fates based on their CO core mass at the end of core helium burning (see Sect.~\ref{sect: methods}). In Fig.~\ref{fig: SN types} we show the final fates in the $M_\mathrm{CO}\,{-}\, M_\mathrm{ini}$ plane. The dark blue region shows models with CO core masses ${<}\,1.03\,\mathrm{M_\odot}$ which could not reach carbon ignition in \textsc{Mesa}, therefore we assume they form CO WD.
With the sky-blue region, we show the CO core masses smaller than $1.37\,\mathrm{M_\odot}$ where ONeMg WDs are expected to form, the green region is for CO core masses between $1.37\,{-}\,1.43\,\mathrm{M_\odot}$ where models explode as ECSN and the pink region for CO core masses ${>}\,1.43\,\mathrm{M_\odot}$ where models form Fe CCSNe \citep[see Sect.~\ref{sect: methods} and ][]{tauris_ultra-stripped_2015}. 

Models with a high convective core overshooting value form larger CO core masses (see Sect.~\ref{sect: core properties}). By increasing the convective core overshooting value, models with lower initial mass form Fe CCSNe. The $7\, \mathrm{M_\odot}$ model with the lowest convective core overshooting forms a CO core mass of $0.9\,\mathrm{M_\odot}$ that becomes a CO WD, while the same initial mass model with the highest convective core overshooting forms a CO core mass of $1.5\,\mathrm{M_\odot}$, massive enough to form a Fe CCSN. For the lowest convective core overshooting value, only the models more massive than $10\,\mathrm{M_\odot}$ can form a Fe CCSN, while for the highest convective core overshooting value the $7\, \mathrm{M_\odot}$ model produces a Fe CCSN.

Using the parametric supernova code of \citet[][see also Sect.~\ref{sect: methods}]{muller_simple_2016} we mark the models that fail to explode but rather collapse into BHs with \hl{black-edged hexagons}. We find black-hole progenitors at CO core masses of ${\sim}\, 7\, \mathrm{M_\odot}$ and ${>}\, 14\, \mathrm{M_\odot}$, which correspond to the first and second compactness peaks. 
For models with the lowest convective core overshooting value, the $24\,\mathrm{M_\odot}$ and the $\hl{40}\, \mathrm{M_\odot}$ models collapse to form BHs. By increasing the convective core overshooting value the models with smaller initial masses form BHs because they have cores of the same compactness as more massive stars with low convective core overshooting. For models with the highest convective core overshooting value, the $19\,\mathrm{M_\odot}$ and the $30\, \mathrm{M_\odot}$ models collapse into BHs. These models correspond to the models with a high compactness parameter (see Sect.~\ref{sect: core properties}).

With blue and red star symbols, we show the models that are expected to produce SN~IIb and SN~Ib/c, respectively. Red supergiant stars by the end of their lives are left with a hydrogen-rich envelope, hence producing SN~IIP. We find that by increasing the convective core overshooting value, the type of SN produced changes (Fig.~\ref{fig: SN types}). Our high-mass models with the highest convective core overshooting value produce SN~IIb and even SN~Ib/c. For example, the $27\, \mathrm{M_\odot}$ model with the highest convective core overshooting produces SN~Ib/c, while the same initial mass model with an intermediate convective core overshooting ($\alpha_\mathrm{ov}\,{=}\,0.30$) produce SN~IIb and those with $\alpha_\mathrm{ov}\,{<}\,0.30$ produce SN~IIP. High-mass models with high convective core overshooting lose most of their hydrogen-rich envelope in stellar winds, hence producing SN~IIb or even SN~Ib/c depending on their envelope mass at the onset of Fe CC (Sect.~\ref{sect: methods}).

\section{Compact remnant masses}\label{sect: Compact-remnants masses}
\begin{figure*}
    \centering
     \includegraphics{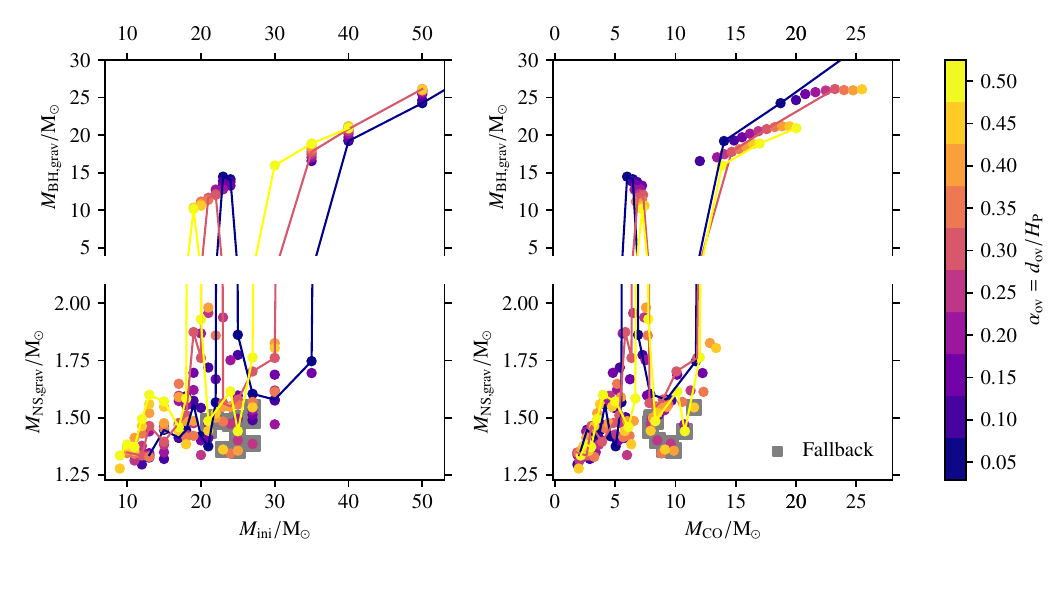}
    \caption{Gravitational neutron-star $M_\mathrm{NS,grav}$ and black-hole masses $M_\mathrm{BH,grav}$ as a function of initial mass (left) and CO core mass (right) for our models up to $M_\mathrm{ini}\,{=}\,50\,\mathrm{M_\odot}$. Different convective core overshooting values are indicated by colors and gray boxes indicate models that are expected to experience fallback.}
    \label{fig: BH- Mini}
\end{figure*}

On the left and right panels of Fig.~\ref{fig: BH- Mini}, we show the gravitational remnant masses of all models that reach Fe CC as a function of the initial mass and as a function of the CO core masses, respectively. 
The remnant mass distribution follows a non-monotonic trend as a function of initial masses and CO core masses similar to the compactness parameter, with the first BH formation peak at $M_\mathrm{ini}\,{\approx}\,19\,{-}\,24\,\mathrm{M_\odot}$ and a second BH formation peak at $M_\mathrm{ini}\,{\approx}\,30\,{-}\,40\,\mathrm{M_\odot}$.

We see that the $24\, \mathrm{M_\odot}$ model with the lowest convective core overshooting value forms a BH of $15\, \mathrm{M_\odot}$. With increasing the convective core overshooting value, the $19\, \mathrm{M_\odot}$ model forms a BH of $10.2\, \mathrm{M_\odot}$. For a higher convective core overshooting value, stars with lower initial masses collapse to form BHs. The same applies to the second peak of BH formation. For models with the lowest convective core overshooting, the $40\, \mathrm{M_\odot}$ model forms a BH of $19.2\,\mathrm{M_\odot}$, and for the highest convective core overshooting the $30\, \mathrm{M_\odot}$ model forms a BH of $15.9\,\mathrm{M_\odot}$. 

For models that experience a direct collapse into a BH, the mass of the remnant depends on the final mass of the progenitor star. The BH remnant masses have a different order on the first and the second BH formation peaks, which is a consequence of the reversed fractional mass loss (see Fig.~\ref{fig: fractional mass loss}). For models in the first BH formation peak ($M_\mathrm{ini}\,{<}\,30\,\mathrm{M_\odot}$), the same initial mass model with a higher convective core overshooting value forms a lower mass BH because they have a smaller final mass. 
For models in the second BH formation peak ($M_\mathrm{ini}\,{>}\,30\,\mathrm{M_\odot}$), the same initial mass model with the highest convective core overshooting forms a more massive BH because they have a higher final mass compared to the same model with a lower convective core overshooting value (see also Fig.~\ref{fig: fractional mass loss}). 
The $40\, \mathrm{M_\odot}$ model with the lowest convective core overshooting forms a BH of $19.2\,\mathrm{M_\odot}$ while the same initial mass model with the highest convective core overshooting forms a BH of $20.9\,\mathrm{M_\odot}$ which correspond to the mass of the progenitor star at the onset of Fe CC.

From our models we predict a minimum BH mass of $10.2\, \mathrm{M_\odot}$ if fallback is not taken into account. No LBV mass loss is applied in our models and we over-predict a maximum BH mass of $43.2\, \mathrm{M_\odot}$ from the model with $M_\mathrm{ini}\,{=}\,70\,\mathrm{M_\odot}$ model with $\alpha_\mathrm{ov}= 0.25$.

We find BH formation for the CO core masses ${\sim}\,7\, \mathrm{M_\odot}$ and ${\geq}\,14\, \mathrm{M_\odot}$ as is shown on the right panel of Fig.~\ref{fig: BH- Mini}, corresponding to the models with a high compactness parameter (see Fig.~\ref{fig: compactness vs Mco}).
For models with CO core masses of ${\sim}\,7\, \mathrm{M_\odot}$ and lowest convective core overshooting value, a BH of $14\,\mathrm{M_\odot}$ is formed, while for the same CO core mass but the highest convective core overshooting we form a BH of $10\,\mathrm{M_\odot}$. The models with the same CO core mass but higher convective core overshooting value produce BHs of lower masses because the progenitor star had a lower mass initially and has a lower final mass at the end of the evolution (see Fig.~\ref{fig: fractional mass loss}). For some of the models, the shock cannot propagate throughout the star to launch a successful explosion and hence they experience a partial fallback of the ejected material (see Sect.~\ref{sect: methods}). 

\begin{figure}
    \centering
     \includegraphics{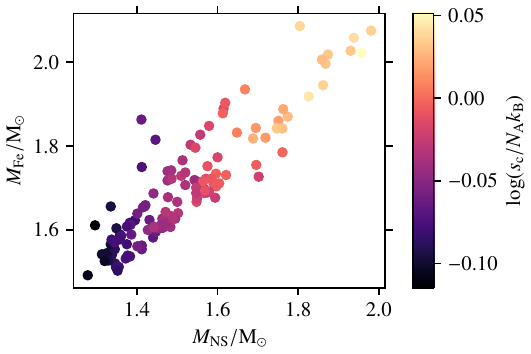}
    \caption{Iron core-mass at the onset of core-collapse as a function of the neutron star mass formed. The colors show the final central specific entropy of the star.}
    \label{fig: iron core mass vs NS mass}
\end{figure}

In Fig.~\ref{fig: iron core mass vs NS mass}, we show the iron core masses $M_\mathrm{Fe}$ at the end of evolution as a function of the NS masses formed for all the models that successfully explode. In colors, we show the final central entropy of the star. The NS masses scale linearly with the iron core masses. We find that we form NS masses in two distinct branches depending on the central entropy of the star. For the same $M_\mathrm{Fe}$ but lower $s_\mathrm{c}$, we consistently form NS of lower masses, compared to the same $M_\mathrm{Fe}$ with higher $s_\mathrm{c}$. The origin of the two branches remains unclear. 

\begin{figure}
    \centering
     \includegraphics{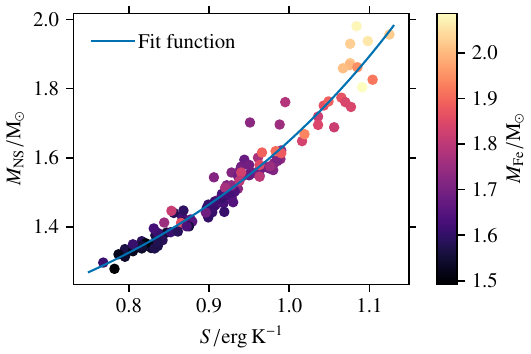}
    \caption{Neutron star mass as a function of the final central specific entropy. The colors indicate the iron-core mass at the onset of Fe CC, and the blue line shows the fit function from Eq.~\ref{eq: entropy-ns}.}
    \label{fig: entropy with fit}
\end{figure}

In Fig.~\ref{fig: entropy with fit}, we show the NS masses as a function of their final central specific entropy with colors showing the iron core masses for all our models that launch an explosion. 
We find a tight relation between the final central specific entropy and the NS mass.
Neutron star masses $M_\mathrm{NS}$ follow an exponential function with the central specific entropy of the star $s_\mathrm{c}$ of the form 
\begin{equation}\label{eq: entropy-ns}
M_\mathrm{NS}=\mathrm{A} \cdot \exp[\mathrm{B}\cdot s_\mathrm{c}]+\mathrm{C}.
\end{equation}
Fitting this function to our NS masses we obtain these fitting parameters $\mathrm{A}=0.048\pm 0.02, \, \mathrm{B}=2.76 \pm 0.46, \,\,\mathrm{and}\,\, \mathrm{C}=0.88 \pm 0.11 $ (uncertainties are $1\sigma$). Knowing this relation, we can infer the central entropy for a given NS mass. 

\section{Supernova explosion properties}\label{sect:  Supernova explosion properties}
\subsection{Correlation to the compactness parameter}

\begin{figure*}
    \centering
     \includegraphics{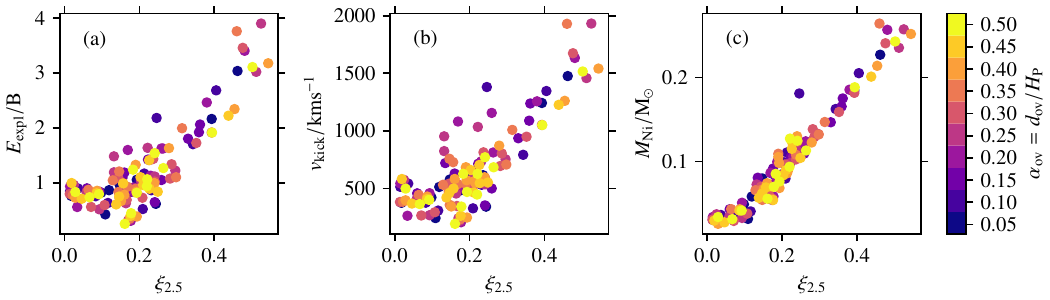}
    \caption{Supernovae explosion properties as a function of the compactness parameter. The explosion energy (a), kick velocity (b), and nickel mass produced (c). The colors indicate the different convective core overshooting values.}
    \label{fig: expl vs compactness}
\end{figure*}

The explosion energy ($E_\mathrm{expl}$), kick velocity ($v_\mathrm{kick}$), and nickel mass production ($M_\mathrm{Ni}$) are calculated in our models using the parametric SN code of \cite{muller_simple_2016} as described in Sect.~\ref{sect:  semi-analytic SN code}. In Fig.~\ref{fig: expl vs compactness}, we show these as a function of the compactness parameter of our models. We find that regardless of the convective core overshooting value, these quantities correlate closely with the compactness parameter. These results are in agreement with the models presented by \cite{schneider_pre-supernova_2021} for single and binary-stripped stars. These correlations are not surprising as they are built into \cite{muller_simple_2016} formalism \hl{and are generally expected for neutrino-driven explosions}. 
\hl{The explosion energy is mostly affected by the accreted mass onto the proto-neutron star, therefore we find a tight correlation with the compactness parameter \citep[][]{schneider_pre-supernova_2021}. The temperature after the shock explosion directly affects the Ni mass production and the kick velocity is directly computed from the explosion energy hence the correlation \citep{schneider_pre-supernova_2021, schneider_bimodal_2023}. We find that the relation with the explosion energy, kick velocity, and Ni mass production is mainly dependent on the final internal structure of the star \citep[compactness or entropy, see also ][]{schneider_pre-supernova_2021,schneider_bimodal_2023} and Fig.~\ref{fig: expl vs compactness}. Any supernova will be affected by this internal structure, as it provides the initial condition for the explosion. 
Different SN models would result in different SN properties, and these correlations are only expected for neutrino-driven SNe. In a magnetically driven engine-like SN model, these correlations would be different.}

The models with a low compactness value show similar values of explosion energy and for $\xi_{2.5}\, {>}\, 0.2 $ the explosion energy increases almost linearly \citep[left panel of Fig.~\ref{fig: expl vs compactness} and Fig.~11 in ][]{schneider_pre-supernova_2021}. 
Typically, the explosion energy of CCSNe is of the order of ${\sim}\, 10^{51}\,\mathrm{erg}\,{=}\, 1\,\mathrm{B}$ \citep{kasen_type_2009}. Analyzing the explosion properties for a population of stars with a fixed convective core overshooting by applying a \citet{salpeter_luminosity_1955} IMF we find median values of $E_\mathrm{expl}$ in the range $0.75\,{-}\,0.98\,\mathrm{B}$ and no trend of $E_\mathrm{expl}$ with convective core overshooting. 

After a successful supernova, the neutron star receives a kick due to asymmetries in the explosion. In the middle panel of Fig.~\ref{fig: expl vs compactness}, we show the kick velocity as a function of the compactness parameter. We find no dependence on the convective core overshooting value, but we see a linear trend with compactness because a higher explosion energy leads to a higher kick velocity \citep{schneider_pre-supernova_2021}. By implying a \cite{salpeter_luminosity_1955} IMF, we find median values of $v_\mathrm{kick}$ for a population of stars with fixed convective core overshooting in the range $420\,{-}\,548\, \mathrm{km\,s^{-1}}$ \citep[not sigma of a Maxwellian distribution, corresponding approximately to $\sigma\,{=}\,265\,\mathrm{km\, s^{-1}}$, ][]{hobbs_statistical_2005}.

The nickel mass produced in the supernova explosion scales linearly with the compactness parameter \citep[see the right panel of Fig.~\ref{fig: expl vs compactness} and Fig.~11 in][]{schneider_pre-supernova_2021}. A higher explosion energy leads to a higher post-shock temperature which then directly affects the Ni mass-produced, hence the correlation.

For models producing SN~IIP, we find a median value of nickel mass production of $M_\mathrm{Ni}\,{\approx}\, 0.066\, \mathrm{M_\odot}$, and for models producing SN~IIb, we find a median value of nickel mass production of $M_\mathrm{Ni}\,{\approx}\,0.101\, \mathrm{M_\odot}$. Most of our models explode as SN~IIP, and only some of them produce SN~IIb and SN~Ib/c (see Fig.~\ref{fig: SN types}). 
For a population of stars with fixed convective core overshooting, we find median values of $M_\mathrm{Ni}$ in the range $0.054\,{-}\,0.077\, \mathrm{M_\odot}$ but no trend of $M_\mathrm{Ni}$ with convective core overshooting. The average inferred observed nickel mass production for SN~IIP is ${\sim}\, 0.032\, \mathrm{M_\odot}$ and for SN~IIb is ${\sim}\, 0.102\, \mathrm{M_\odot}$ \citep{anderson_meta-analysis_2019}.

\subsection{SN ejecta mass and SN~IIP lightcurves} 
\begin{figure}
    \centering
     \includegraphics{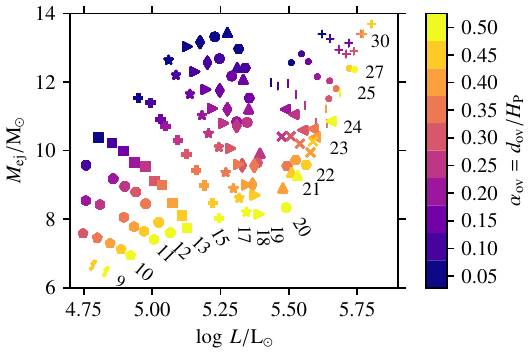} 
    \caption{Ejecta mass of our set of models expected to produce a SN explosion as a function of their luminosity at the onset of CC. Markers indicate models with the same initial mass, and numbers are initial masses in solar masses. Different colors show different convective core overshooting values.}
    \label{fig: ejecta mass vs luminosity}
\end{figure}

In Fig.~\ref{fig: ejecta mass vs luminosity}, we show the SN ejecta mass as a function of the luminosity at the onset of Fe CC for our models that are expected to successfully explode. The SN ejecta mass follows a pattern with initial masses and the luminosity at the onset of Fe CC for different values of convective core overshooting (see Fig.~\ref{fig: ejecta mass vs luminosity}). This applies to the initial mass models ${<}\,20\, \mathrm{M_\odot}$ and $\log(L/\mathrm{L_\odot})\,{<}\,5.4$. Models with the same initial mass but higher convective core overshooting reach a higher luminosity at the onset of Fe CC and have a smaller ejecta mass compared to the same initial mass model with lower convective core overshooting. This is because the low-mass models ($M_\mathrm{ini}\,{<}\,20\, \mathrm{M_\odot}$) with high convective core overshooting lose more mass during the evolution and are left with a smaller envelope that can be ejected during the SN explosion compared to the same initial mass models with a smaller convective core overshooting value (Fig.~\ref{fig: fractional mass loss}). In contrast, for models with higher initial $M_\mathrm{init}\,{>}\, 20\, \mathrm{M_\odot}$ and $\log(L/\mathrm{L_\odot})\,{>}\,5.4$, there is no longer a clear trend in the ejecta mass. Because these models correspond to the ones where the fractional mass loss reaches a turning point (see also Fig.~\ref{fig: fractional mass loss}). Almost all of these models lose a significant fraction of their initial mass due to winds, affecting the final ejected mass.

\begin{figure}
    \centering
     \includegraphics{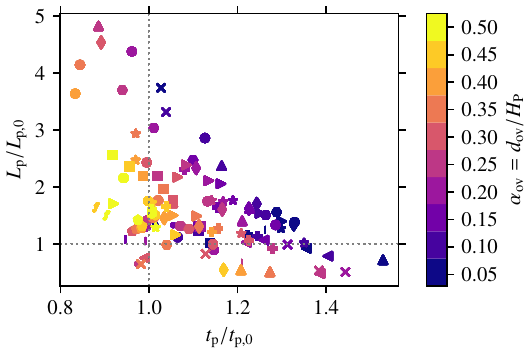}
    \caption{Expected plateau luminosity of SN light-curve as a function of the duration of the plateau for our SN~IIP progenitor models. Markers indicate models of the same initial mass, and colors indicate different convective core overshooting values. The black-dotted lines indicate typical values of  $L_\mathrm{p,0}\,{=}\,10^{42}\,\mathrm{erg/s}$ and $t_\mathrm{p,0}\,{=}\,100\,\mathrm{days}$ that are calibrated for RSGs with an $M_\mathrm{ej}\,=\,10\,\mathrm{M_\odot}$, $E_\mathrm{expl}\,=\,10^{51}\,\mathrm{erg}$ and $R\,=\,500\,\mathrm{R_\odot}$.}
    \label{fig: plateau luminosity}
\end{figure}

For all the models that are expected to produce SN~IIP, we compute the plateau luminosity and the duration of the plateau of the SN light curve as described in Sect.~\ref{sect:  semi-analytic SN code}. 
We calibrate the scaling relations to $L_\mathrm{p,0}$ and $t_\mathrm{p,0}$ for a typically assumed ejecta mass $M_\mathrm{ej}\,{\hl{=}}\,10\,\mathrm{M_\odot}$, explosion energy $E_\mathrm{expl}\,{\hl{=}}\,10^{51}\,\mathrm{erg}$ and progenitor radius $R\,{\hl{=}}\,500\,\mathrm{R_\odot}$.
The values are expressed in units of plateau luminosity $L_\mathrm{p,0}\,{=}\,10^{42}\,\mathrm{erg/s}$ and duration of plateau $t_\mathrm{p,0}\,{=}\,100\,\mathrm{days}$ which are typical values for SN~IIP \citep{popov_analytical_1993}.
Models with a higher convective core overshooting value tend to have a shorter duration of the plateau luminosity (Fig.~\ref{fig: plateau luminosity}). Since the duration of the plateau is proportional to the square root of the ejecta mass (Eq.~\ref{eq: plateau t}), the models with high convective core overshooting and small ejecta material have the shortest plateau duration. The plateau luminosity is given by a combination of the explosion energy, the ejected mass, and the radius of the progenitor star (see Eq.~\ref{eq: plateau L}). For increasing convective core overshooting values the ejecta mass generally decreases, and the radius of the progenitor star is smaller, these two have an opposite effect on the plateau luminosity. However, the impact of the radius seems to dominate the ejecta mass of our models, hence overall we see a lower plateau luminosity for models with the highest convective core overshooting (see Fig.~\ref{fig: plateau luminosity}).

\section{Explosion sites in HRD and missing RSG problem}\label{sect: HRD and missing RSG problem}

\begin{figure*}
    \centering
     \includegraphics{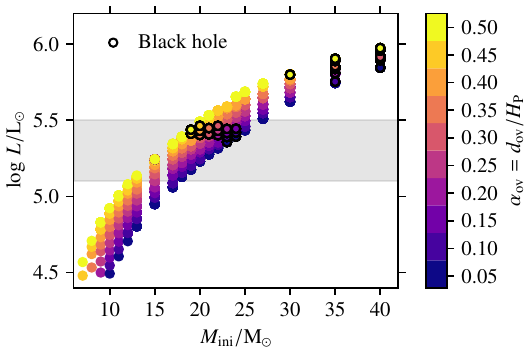}
     \includegraphics{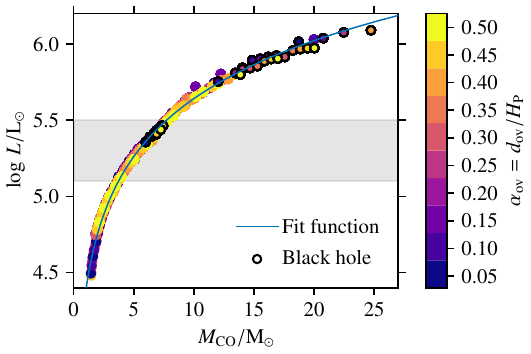}
    \caption{The luminosity of pre-SN models at the end of core-carbon burning, as a function of initial mass (left) and CO core mass (right). The colors represent different values of convective core overshooting and the black circles indicate models that are expected to collapse into BHs. The gray-shaded region is the range of luminosity in which no SN~IIP progenitors have been observed so far $5.1 \,{\leq}\, \log(L/\mathrm{L_\odot})\,{\leq}\, 5.5$ \citep{davies_red_2020}. The blue line in the right panel shows our fit function to the luminosity and CO core masses of our models (see text).}
    \label{fig: luminosity-M_ini}
\end{figure*}

In Fig.~\ref{fig: luminosity-M_ini}, we show the luminosity at the end of core carbon burning as a function of the initial mass (left panel) and as a function of the CO core mass (right panel) for all our models that are expected to produce Fe CCSNe. Models with a higher convective core overshooting value have higher CO core masses and higher luminosities (see Sect.~\ref{sect: pre-sn evolution}). 
Stars of initial masses between $19\,\mathrm{M_\odot}$ to approximately $24\,\mathrm{M_\odot}$ and those of $M_\mathrm{ini}\,{\geq}\,30\,\mathrm{M_\odot}$ collapse to form BHs (see Sect.~\ref{sect: Compact-remnants masses}). We find that BH progenitors have similar luminosities at the end of core carbon burning. The low-mass BH progenitor models have luminosities of $5.35\,{\leq}\,\log(L/\mathrm{L_\odot})\,{\leq}\, 5.47$ (see the left panel of Fig.~\ref{fig: luminosity-M_ini}). The high-mass BH progenitors have luminosities of $\log(L/\mathrm{L_\odot})\,{\geq}\,5.8$ corresponding to the second peak of BH formation (left panel of Fig.~\ref{fig: luminosity-M_ini}). 

As the right panel of Fig.~\ref{fig: luminosity-M_ini} shows, the luminosity at the end of core carbon burning is essentially only a function of the CO core mass. The luminosity of red supergiants can be described by the core mass-luminosity relation. We fit a power law function of the form
 \begin{equation}\label{eq: CO_mass-L relation}
    \log(L/\mathrm{L_\odot})\,{=}\,A\,{+}\,B \,{\cdot}\, \log(M_\mathrm{CO}/\mathrm{M_\odot}),
\end{equation} 
with these fitting parameters $A\,{=}\,4.372\,{\pm}\, 0.005$ and $B\,{=}\,1.268\,{\pm}\, 0.006$ and $1\,\sigma$ uncertainties. 

Using the fit function in Eq.~\ref{eq: CO_mass-L relation}, we can infer the CO core mass of the observed pre-SN luminosities of progenitor stars and even predict the final fate. 
In many studies \citep{smartt_progenitors_2009,smartt_observational_2015, davies_initial_2018} the luminosity of SN progenitors is used to infer their initial mass, but as we show in Fig.~\ref{fig: luminosity-M_ini} there is no clear relation between $\log(L/\mathrm{L_\odot})$ and $M_\mathrm{ini}$.
For example, for the luminosity of $\log(L/\mathrm{L_\odot})\,{=}\,5.1$ the progenitor star could be a $13\,\mathrm{M_\odot}$ star with the highest convective core overshooting or a $17\,\mathrm{M_\odot}$ star with the lowest convective core overshooting. Instead, we can infer their CO core mass more precisely. For example for the SN progenitor with a luminosity of $\log(L/\mathrm{L_\odot})\,{=}\,5.1$ using the fit function we infer a CO core mass of $3.7\,\mathrm{M_\odot}$. 

The gray-shaded area in Fig.~\ref{fig: luminosity-M_ini} indicates the missing red supergiant region \ie that there are no observed SN~IIP progenitor stars for $\log(L/\mathrm{L_\odot})\,{>}\,5.1$, which may be in tension with the upper luminosity of red supergiants of $\log(L/\mathrm{L_\odot})\,{=}\,5.5$ \citep{davies_red_2020}. 
A suggested solution for the missing progenitors of SN~IIP is that RSGs more luminous than $\log(L/\mathrm{L_\odot})\,{>}\,5.1$ collapse into BHs \citep{smartt_observational_2015}. However, our simulations predict a successful explosion up to $\log(L/\mathrm{L_\odot})\,{=}\,5.35$ and BH formation in a much narrower range of luminosities of $5.35 \,{\leq}\, \log(L/\mathrm{L_\odot})\,{\leq}\, 5.47$.

\begin{figure*}
    \centering
     \includegraphics{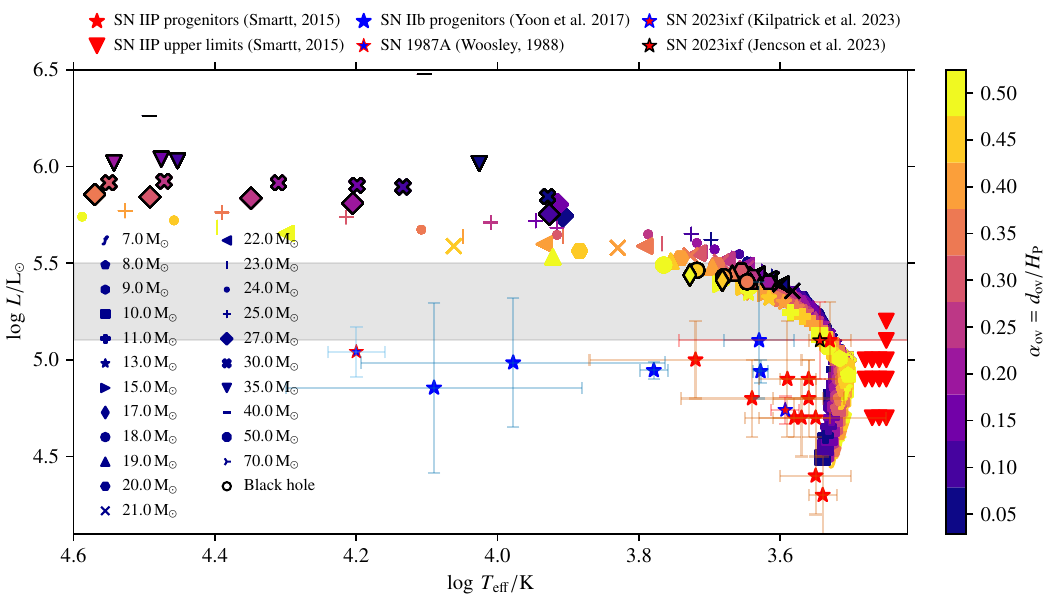}
    \caption{Hertzsprung-Russell diagram of all our calculated models at the end of core-carbon burning compared to observed supernova progenitors.~Symbols show initial masses, and colors indicate different convective core overshooting values.~Symbols with black edges are models that are expected to collapse into BHs. As in Fig.~\ref{fig: luminosity-M_ini} the grey shaded region shows the missing RSG region.~Red star symbols represent the observed SN~IIP progenitor stars, with uncertainties in $\log(L/\mathrm{L_\odot})$ and $\log (T_\mathrm{eff}/\mathrm{K})$ from \cite{smartt_observational_2015}. The red triangles show the upper limits in the luminosity of the observed SN~IIP progenitors from \cite{smartt_observational_2015} for which effective temperatures are not known. The red star symbol with blue and black edges shows the SN 2023ixf progenitor from \cite{kilpatrick_sn2023ixf_2023} and \cite{jencson_luminous_2023} respectively. The blue star symbols indicate the observed progenitor stars of type SN~IIb as compiled in \cite{yoon_type_2017}. The blue star symbol with red edges indicates the SN 1987A progenitor by \cite{woosley_sn_1988}.}
    \label{fig: HRD with SN progenitors}
\end{figure*}

In Fig.~\ref{fig: HRD with SN progenitors}, we show an HRD of the entire set of our models at the end of core-carbon burning. The black-edged markers show the models which do not produce a supernova but rather collapse into BHs based on the results from using the parametric SN code of \cite{muller_simple_2016}.
We compare our models with the observed progenitors of SN~IIP from \citet{smartt_observational_2015}. The progenitors of SN 2009kr and SN 2009md are removed from the \cite{smartt_observational_2015} sample following \cite{maund_whatever_2015}. For completeness, we also show the SN 1987A progenitor \citep[][]{woosley_sn_1988}.

We find a tension between our models and the observed SN~IIP progenitors in both luminosity and effective temperature. However, it should be noted that our models are not calibrated to the effective temperatures of RSGs, and the uncertainties in observed SN~IIP progenitors are quite large. For example, the uncertainty in the luminosity of the observed SN~IIP progenitors extends up to a luminosity of $\log(L/\mathrm{L_\odot})\,{\approx}\,5.3$. 
The upper limits in the luminosity of the SN~IIP progenitors have values up to $\log(L/\mathrm{L_\odot})\,{=}\, 5.2$ which is in the region of the ``missing RSG problem''. 
Our models cannot explain the two least luminous SN~IIP progenitors, because they are too faint. These are the progenitors of SN2003gd and SN2005cs with luminosities of $\log(L/\mathrm{L_\odot})\,{=}\, 4.3$ and $\log(L/\mathrm{L_\odot})\,{=}\, 4.4$, respectively.
Using the luminosity-CO core mass relation from Eq.~\ref{eq: CO_mass-L relation} we infer their CO core mass and we find values of ${\sim}\,0.88\,\mathrm{M_\odot}$ and ${\sim}\,1.04\,\mathrm{M_\odot}$ for SN2003gd and SN2005cs, respectively. For such CO core masses, we would predict WD formation. 
This discrepancy could be resolved if the observed luminosity of SN~IIP progenitors were underestimated (see also Sect.~\ref{sect: missing RSG problem} for a more detailed discussion).  

Additional to the already known progenitors of type IIP SNe from \cite{smartt_observational_2015} the progenitor star of SN 2023ixf\footnote{ SN2023ixf occurred during the writing of this paper and the progenitor's properties are not clear yet with different values on $T_\mathrm{eff}$ and $\log(L/\mathrm{L_\odot})$ being reported.} has recently been observed and its progenitor star appears to be a red supergiant that exploded as a type II SN \citep{kilpatrick_sn2023ixf_2023, jencson_luminous_2023}. Different groups report different progenitor star properties (see Tab.~\ref{tab: SN 2023}). However, the SN 2023ixf progenitor star may be one of the most luminous progenitors ever detected \citep{jencson_luminous_2023}. We show the progenitor star of SN 2023ixf from \cite{kilpatrick_sn2023ixf_2023} and \cite{jencson_luminous_2023} in Fig.~\ref{fig: HRD with SN progenitors} by red star markers with blue and black edges respectively. According to \cite{jencson_luminous_2023} the progenitor's luminosity is $\log(L/\mathrm{L_\odot})\,{=}\,5.1 \pm 0.2$ which falls in the missing red supergiant regime but it still lies in the region where we predict a successful SNe with luminosity up to $\log(L/\mathrm{L_\odot}) \,{\approx}\, 5.3$.

\begin{table}[h]
    \centering
    \caption{Luminosity and effective temperature of SN 2023ixf progenitor star reported by two different groups.} \label{tab: SN 2023}
    \begin{tabular}{|c|c|c|}
    \hline
 $\log(L/\mathrm{L_\odot})$ & $T_\mathrm{eff}/\mathrm{K}$  & Reference\\
\hline
  $4.74 \pm 0.07$ & $3920^{+200}_{-160}$ & \cite{kilpatrick_sn2023ixf_2023}\\
                          & &  \\
 $ 5.1 \pm 0.2$  & $3500^{+800 }_{-1400}$ & \cite{jencson_luminous_2023}\\
 \hline
    \end{tabular}

\end{table}

SN~IIb progenitors from \citet{yoon_type_2017} are indicated with blue star symbols in Fig.~\ref{fig: HRD with SN progenitors}. Some of our models form SN~IIb (see Fig.~\ref{fig: SN types}), and comparing them to the observed SN~IIb progenitors and the SN 1987A progenitor we find an offset of ${\geq}\,0.5\,\mathrm{dex}$ in luminosity. Our single-star models cannot explain these SN progenitors. This may be expected, given that the progenitor star of SN 1987A is likely a product of binary-star evolution, stellar merger \citep{podsiadlowski_progenitor_1992}. Similarly, as studied by \cite{yoon_type_2017}, the progenitors of SN~IIb most likely are the result of donor stars in binaries that transferred their outer envelope during Case B binary mass transfer.

\section{Discussion}\label{Discussion}

\subsection{Uncertainties in stellar winds} \label{sect: winds}
Wind mass loss rates are not fully understood and it is difficult to model them analytically \citep{smith_mass_2014}. Winds influence the evolution of stars, change their internal structure and affect their final fate \citep[\eg][]{chiosi_evolution_1986, renzo_systematic_2017}.

For the same initial mass model but different wind mass loss efficiency or mass loss algorithm, the compactness parameter can change significantly \citep[][]{renzo_systematic_2017}. Stronger wind mass loss leads to stars losing more of their envelope and having a higher carbon abundance in the CO core at the end of core helium burning, similar to what we find for our high initial mass models with the highest convective core overshooting. In the most extreme cases, winds remove the entire hydrogen-rich envelope such that stars evolve like binary-stripped stars \cite[\cf][]{schneider_pre-supernova_2021, laplace_different_2021}. 
This also affects the compactness, and thus the explodability of these stars  \citep{patton_towards_2020, schneider_pre-supernova_2021,laplace_different_2021}. Enhanced wind mass-loss rates affect the final mass of the progenitor stars, hence affecting the remnant masses left behind. Higher wind mass-loss rates lead to reduced CO-core and thus iron core masses. Hence, the NSs formed will have smaller masses. Depending on how strong the mass loss rates are, they can change the final outcome, hence changing the number of NSs produced. As shown in Sect.~\ref{sect: Compact-remnants masses}, the BH remnant masses are linked to the final mass of the progenitor star. The same initial mass models but with an increased mass-loss rate would form a smaller mass BH. 
For stronger mass-loss rates the progenitor models will have a smaller envelope mass, and BHs formed by fallback would also have smaller masses.

Models with enhanced wind mass-loss rates would lose more of their envelope mass, hence producing hydrogen-poor SN explosions \citep{smith_mass_2014, renzo_systematic_2017} with smaller ejecta masses.
A smaller ejected mass for SN~IIP shortens the duration of the plateau luminosity, similar to what we found for models with an increased convective core overshooting value in Sect.~\ref{sect:  Supernova explosion properties}. 
An enhanced wind mass-loss rate after core-helium burning would not affect the internal structure of the star or the explodability, unless helium or carbon-rich material is lost in stellar winds, in such case our models would behave like binary-stripped stars \citep{schneider_pre-supernova_2021, laplace_different_2021}.

The opposite effect is expected for reduced mass-loss rates, stars would have a larger envelope mass and consequently a smaller carbon abundance in the core after core-helium burning. Thus, we expect them to be on average more difficult to explode.
Since their progenitor models would have higher final masses, the BH remnants would be more massive. In the case of a successful SN explosion, for reduced mass loss rates, models are more likely to produce SN~IIP and have a larger ejecta mass.

Some massive stars may experience LBV-like mass loss and eruptions leading to a sudden increase in luminosity and extended mass loss \citep{smith_mass_2014}. Our models do not take this into account. However, if these instabilities occur before helium depletion, they can change the internal structure of stars by removing the entire hydrogen-rich envelope. The internal structure of these models would be similar to Case-B binary-stripped stars \citep[\cf][]{schneider_pre-supernova_2021, laplace_different_2021}, consistently having a larger carbon abundance in the core and the compactness peak shifted to higher CO core masses. Hence, these models explode more easily and form NSs \citep{schneider_pre-supernova_2021}. As reported by \cite{schneider_pre-supernova_2021} for the same CO core mass, the remnant masses of BHs are smaller for models that go through Case B mass transfer, which also applies to our high mass models if we were to include the eruptive mass loss in our simulations.

\subsection{Convective boundary mixing}\label{sect: overshooting}

In our stellar models, we have assumed step convective core overshooting above convective-core hydrogen and helium burning. By introducing the step convective core overshooting in our calculation, we assume an instantaneous mix of chemical elements to a distance $d_\mathrm{ov}$ above the convective boundary. The size of the convective core overshooting layer is uncertain and is not well constrained by observations \cite[][]{schroder_critical_1997, stancliffe_confronting_2015, salaris_chemical_2017}. Previous studies have shown that the convective core overshooting parameter is mass dependent and that higher-mass stars have a larger convective core overshooting parameter \citep[\cf][]{schroder_critical_1997, higl_calibrating_2021, jermyn_convective_2022, ahlborn_stellar_2022}. At present, the exact value of the convective core overshooting for different initial masses is not yet parametrized. Knowing the mass dependence of the convective core overshooting parameter, one could perform population synthesis simulations to more accurately predict the luminosity of the SN and BH progenitors. Because the luminosity of the BH progenitors is not influenced by the initial mass of the models but only the CO core mass, it would not change the luminosity region where we predict BH formation (\cf Fig.~\ref{fig: luminosity-M_ini}).

Because of the steep gradient of the chemical composition above the convective helium-burning core, the convective boundary mixing may be smaller during this phase \citep{langer_presupernova_2012, jermyn_convective_2022}. Less convective boundary mixing above the convective-helium burning core would lead to a smaller CO core, and also to a higher carbon abundance in the core after helium burning. This would translate to the compactness peak shifting to lower CO core masses, affecting the explodability of stars. For a compactness peak at smaller CO core masses the BH progenitors would have a slightly lower luminosity. The convective-core overshooting on the later burning stages would have the same impact as assuming an increased convective core overshooting value, leading to higher final core masses. However, convective core overshooting \hl{in} the later burning stages may induce shell mergers \citep[\cf][]{mocak_turbulent_2018, andrassy_3d_2020, yadav_large-scale_2020}, hence influencing the explodability \citep{oconnor_black_2011,sukhbold_compactness_2014, schneider_pre-supernova_2021}, as well as changing the signatures of the SN explosions \citep{andrassy_3d_2020}.

\subsection{Rotationally-induced mixing}\label{convection vs rotation}

Convective core overshooting causes additional mixing of the fresh fuel into the convective burning core, increasing the mass of the core and the luminosity of the star. Similar results are found for rotating stars because of rotationally-induced mixing \citep[\cf][]{heger_presupernova_2000, maeder_evolution_2000, maeder_rotating_2012, hirschi_stellar_2004, brott_rotating_2011, nguyen_parsec_2022}. Rotationally-induced mixing has a strong impact on the pre-SN structure and on the type of SN produced \citep{hirschi_stellar_2004}. They found that rotation changes the internal structure of stars with initial masses up to $30\,\mathrm{M_\odot}$. However, for models more massive than $30\,\mathrm{M_\odot}$ the rotation does not play a significant role, since their evolution is mostly dominated by the effects of stellar winds.
They show that the rotation influences the carbon abundance in the core, in such a way that highly rotating models have a smaller carbon abundance in the core, but that changes for the most massive stars where the impact of winds dominates that of rotation, similarly to what we found for our models with an increased convective core overshooting value (\cf Fig.~\ref{fig: carbon mass fraction}).

Studies from \cite{fichtner_mechanical_2022} show that for non\hl{-}rotating models with initial masses $M_\mathrm{ini}\,{\leq}\,25\,\mathrm{M_\odot}$, the mass loss is dominated by RSG winds, and stars with $M_\mathrm{ini}\,{\geq}\,25\,\mathrm{M_\odot}$ experience BSG/LBV and WR winds (see their Fig.~A1).
For the non-rotating models, they find a maximum fractional mass loss at the $M_\mathrm{ini}\,{=}\,25\,\mathrm{M_\odot}$. In highly rotating models, this maximum is shifted to lower initial masses, because an increase in rotation makes models stay longer on the RSG branch and lose the highest fraction of their mass. More massive stars become blue supergiants and are subject to the WR winds \citep{fichtner_mechanical_2022}. 
Similar to what we find in our models, with increasing convective core overshooting value, models with the lower initial mass lost the highest fraction of their mass while the more massive ones evolve\hl{d} to become WR stars where mass-loss rates are weaker.

\subsection{Missing RSG problem}\label{sect: missing RSG problem}

We compared observed SN~IIP progenitors with our models and found that there is an offset in effective temperature between observations and our models. This is not so surprising given that: (a) the inferred effective temperature from observations of IIP SN progenitors is highly uncertain because of the lack of high-quality data, the effect of interstellar and circumstellar extinction, and the systematic uncertainties in the methods used to derive effective temperatures \citep[\eg][]{davies_temperatures_2013}, (b) our models are not calibrated to reproduce the observed effective temperature of RSGs.

The problem of missing SN~IIP progenitors of luminosities of $\log(L/\mathrm{L_\odot})\,{=}\, 5.1\,{-}\,5.5$ has been investigated extensively. One interpretation is that the red supergiants above $\log(L/\mathrm{L_\odot})\,{=}\, 5.1$ collapse to form BHs \citep[][]{smartt_observational_2015}. However, our models predict successful supernovae above this limit and black-hole formation for a much narrower range of luminosities $5.35 \,{\leq}\, \log(L/\mathrm{L_\odot})\,{\leq}\, 5.47$ as seen in Fig.~\ref{fig: luminosity-M_ini}. 
Our range of BH formation is in agreement with the one observed disappearing red supergiant star N6946\, BH1\footnote{\hl{\cite{beasor_jwst_2023} claim that the BH candidate shows infrared luminosity, suggesting that the star may not have collapsed, but rather shed some envelope and is now dark in bands available before JWST. This paper appeared during the refereeing stage and if the star has indeed not collapsed one might have to interpret this source differently.}}. Its progenitor star had a luminosity of $\log(L/\mathrm{L_\odot})\,{\approx}\, 5.3\,{-}\,5.5$ \citep{gerke_search_2015, kochanek_survey_2008, adams_search_2017}. This seems to confirm our prediction for the BH formation region. Even after changing one of the most uncertain parameters of stellar evolution, the convective core overshooting parameter, the missing red-supergiant problem remains but in a smaller range of luminosity.

\hl{Another} possible explanation for the absence of RSGs of $\log(L/\mathrm{L_\odot})\,{\geq}\, 5.1$ could be that these stars lose a large amount of mass during their evolution, and are progenitors of SN~IIb. Hydrodynamic instabilities of massive stars during the late burning stages may lead to eruptive mass loss \citep{yoon_evolution_2010, smith_preparing_2014}. This means that SN progenitor stars may have dust surrounding them, which could make the star appear fainter. 
For example, the observed luminosities of SN2003gd and SN2005cs are so faint that we infer $M_\mathrm{CO}$ of $0.88\,\mathrm{M_\odot}$ and $1.04\,\mathrm{M_\odot}$ from our fit in Eq.~\ref{eq: CO_mass-L relation}, respectively. Stars with such low $M_\mathrm{CO}$ are not expected to lead to SNe but should form WD instead.
There seems to be a need for better constraining the luminosities of the observed SN progenitors \citep[see also][]{davies_initial_2018}.

We find that the pre-SN luminosities cannot be directly related to the $M_\mathrm{ini}$. For example, for the SN progenitors with a luminosity of $\log(L/\mathrm{L_\odot})\,{=}\, 5.1$, its progenitor star could be of the initial mass of range $13\,{-}\,17\,\mathrm{M_\odot}$ with different convective core overshooting value. Instead, as a function of the CO core mass, they form the same CO core mass ${\approx}\,3.5\,\mathrm{M_\odot}$. Therefore, we urge the community to infer the CO core masses rather than the initial masses from the observed luminosity of SN progenitors. 


\section{Conclusions}\label{Conclusion}
We study the impact of convective-core overshooting on the evolution and final fate of single massive stars. Using the 1D stellar evolution code \textsc{Mesa}, we evolve stars in the mass range of $5\,{-}\,70\, \mathrm{M_\odot}$ for different step convective core overshooting values of $0.05\,{-}\,0.50\,H_\mathrm{P}$.
The pre-supernova stellar models are analyzed using the parametric supernovae code of \cite{muller_simple_2016}. 
 
We find that models with different convective core overshooting values have different evolutionary tracks, and the changes persist until the end of their lives. Also, the explodability, explosion properties, and BH and NS masses are affected by convective core overshooting. We compare our pre-SN models to the observed progenitors of type IIP and IIb SN.\\
Our main findings can be summarized as follows.

\begin{itemize}
    
    \item We find that convective core overshooting affects the final masses of stars differently depending on their initial masses and mass loss history. Models with initial masses of $ {\leq}\, 20\,\mathrm{M_\odot}$ and the highest convective core overshooting have a higher fractional mass-loss compared to the same initial mass model with the lowest convective core overshooting. For initial masses ${\geq}\,30\,\mathrm{M_\odot}$, this trend is reversed and models with the highest convective core overshooting have a smaller fractional mass-loss. Moreover, the factional mass loss reaches a  maximum of 50\% in stars of initial masses of $20\,{-}\,35\,\mathrm{M_\odot}$ depending on the convective core overshooting value.
    
    \item \textit{The explodability and final fate of stars are mostly dominated by the CO core mass $M_\mathrm{CO}$ and the central carbon abundance $X_\mathrm{C}$ at the end of core helium burning}. Convective core overshooting mainly affects $M_\mathrm{CO}$ and less so $X_\mathrm{C}$. Models up to $13\,\mathrm{M_\odot}$ with the highest convective core overshooting value have a slightly smaller $X_\mathrm{C}$ at the end of core-helium burning compared to the same model with the lowest convective core overshooting. 
    The most massive stars (${\geq}\,30\,\mathrm{M_\odot}$) models with the highest convective core overshooting value tend to have higher $X_\mathrm{C}$ in the core. Models with initial masses between $13\,{-}\,30 \,\mathrm{M_\odot}$ show no clear trend of $X_\mathrm{C}$ with convective core overshooting parameter.
    
    \item For models with the lowest convective core overshooting, we find compactness peak at initial masses of ${\sim}\,24\,\mathrm{M_\odot}$ and high compactness for $M_\mathrm{ini}\,{\geq}\,40\,\mathrm{M_\odot}$. By increasing the convective core overshooting value, the compactness landscape is shifted towards lower initial masses by $5\,\mathrm{M_\odot}$ and $10\,\mathrm{M_\odot}$ for the first and second peak, respectively. As a function of the CO core mass, we find the compactness peaks at CO core masses of ${\approx}\,7\,\mathrm{M_\odot}$ and $14\,\mathrm{M_\odot}$. Increasing the convective core overshooting value shifts the first compactness peak toward higher CO core masses by $1.3\,\mathrm{M_\odot}$, while the second peak of compactness stays constant.

    \item The compactness parameter, central specific entropy, iron core mass and the binding energy above the iron core all follow the same trend with $M_\mathrm{CO}$. No matter which criterion for explodability we use the models at $M_\mathrm{CO}\,{\approx}\,7\,\mathrm{M_\odot}$ are more difficult to explode, and they are expected to collapse into BHs.

    \item The lowest BH mass is $15\,\mathrm{M_\odot}$ for the models with the lowest convective core overshooting value, whereas it is $10.2\,\mathrm{M_\odot}$ for the models with the highest convective core overshooting value. In the first BH formation peak, models with larger convective core overshooting tend to form lower-mass BHs compared to those with smaller convective core overshooting. In the second BH formation region, this order is reversed and models with the same initial mass but a larger convective core overshooting tend to form higher-mass BHs. This is a direct consequence of the different wind mass loss regimes.
    
    \item We find a close relation of the NS masses with the central specific entropy of the star at the core-collapse.

    \item In our models the explosion energy $E_\mathrm{expl}$, kick velocity $v_\mathrm{kick}$, and nickel mass production $M_\mathrm{Ni}$ follow almost linear trends with the compactness parameter, and we find no correlation with the convective core overshooting value.

    \item The luminosity and the duration of the plateau of SN~IIP lightcurves depend on the radius and ejecta mass of SN progenitors and thus on convective core overshooting. Our highest convective core overshooting models tend to have a shorter plateau duration and less luminous lightcurves.
    
     \item The luminosities $\log(L/\mathrm{L_\odot})$ of our SN progenitors are solely determined by the CO core mass. We find that BH formation occurs for progenitors with luminosities of $5.35\,{\leq}\,\log(L/\mathrm{L_\odot})\,{\leq}\, 5.47$ from stars with $M_\mathrm{CO}\,{\approx}\,7\,\mathrm{M_\odot}$ and at $\log(L/\mathrm{L_\odot})\,{\geq}\,5.8$ from stars with $M_\mathrm{CO}\,{\geq}\,14\,\mathrm{M_\odot}$ independently of the convective core overshooting value. The former BH formation region is in agreement with the observation of the disappearing star N6946\,BH1. 
    
    \item Because convective boundary mixing is still very uncertain, we show that one cannot easily relate $\log(L/\mathrm{L_\odot})$ of SN~IIP progenitors to their initial mass. Instead, one can infer $M_\mathrm{CO}$ quite robustly.
\end{itemize}
  
We conclude that convective core overshooting cannot solve the missing red supergiant problem, i.e. a lack of detected SN~IIP progenitors with $\log(L/\mathrm{L_\odot})\,{=}\,5.1-5.5$, but our models demonstrate that some stars in this luminosity range indeed collapse to BHs.

\begin{acknowledgements}
The authors acknowledge support from the Klaus Tschira Foundation.
This work has received funding from the European Research Council (ERC) under the European Union’s Horizon 2020 research and innovation program (Grant agreement No.\ 945806) and is supported by the Deutsche Forschungsgemeinschaft (DFG, German Research Foundation) under Germany’s Excellence Strategy EXC 2181/1-390900948 (the Heidelberg STRUCTURES Excellence Cluster).
\end{acknowledgements}

\bibliographystyle{aa}
\bibliography{references}

\newpage
\begin{appendix}
\section{Table with the properties of our models}\label{appendix}
\include{Table.tex}

\end{appendix}

\end{document}

%% file: Table.tex
\onecolumn
\begin{landscape}
\tabcolsep=0.15cm
    \renewcommand{\LTcapwidth}{\linewidth}
    \begin{longtable}{c c c c c c c c c c c c c c c c c c c c }
        \caption{\label{tab:models}Properties of all our models used throughout the paper. Explosion properties of models that were not computed to core-collapse are empty. We give the initial masses  $M_\mathrm{ini}$, convective core overshooting values $\alpha_\mathrm{ov}$, helium-core masses $M_\mathrm{He}$, CO-core masses $M_\mathrm{CO}$, iron-core masses $M_\mathrm{Fe}$, carbon abundance at the end of core-helium burning $X_\mathrm{C}$, final mass of the models $M_\mathrm{final}$, compactness value $\xi_\mathrm{2.5}$, central specific entropy at the onset of iron-core collapse $s_\mathrm{c}$, the gravitational binding energy of the material above the iron core $-E_\mathrm{bind}$, the predicted final fate, the gravitational remnant masses $M_\mathrm{rm}$, explosion energy $E_\mathrm{expl}$, kick velocity $v_\mathrm{kick}$, nickel mass $M_\mathrm{Ni}$, ejecta mass $M_\mathrm{ej}$, plateau luminosity $L_\mathrm{p}/\mathrm{L_{p,0}}$, duration of the plateau $t_\mathrm{p}/\mathrm{t_{p,0}}$, luminosity at the end of core-helium burning $\log L_\mathrm{c}$ and the effective temperature at  the end of core-helium burning $\log T_\mathrm{eff}$ }.\\
        \toprule
        $M_\mathrm{ini}$ & $\alpha_\mathrm{ov}$ & $M_\mathrm{He}$ & $M_\mathrm{CO}$  & $M_\mathrm{Fe}$  & $X_\mathrm{C}$  & $M_\mathrm{final}$  & $\xi_\mathrm{2.5}$ & $s_\mathrm{c}$ & $-E_\mathrm{bind}$ & Final fate? & $M_\mathrm{rm}$ & $E_\mathrm{expl}$ & $v_\mathrm{kick}$ & $M_\mathrm{Ni}$ & $M_\mathrm{ej}$  & $L_\mathrm{p}/\mathrm{L_{p,0}}$  & $t_\mathrm{p}/\mathrm{t_{p,0}}$  & $\log L_\mathrm{c}$  & $\log T_\mathrm{eff}$  \\
        
        ($\mathrm{M_\odot}$) & ($d_\mathrm{ov}/\mathrm{H_p}$) & ($\mathrm{M_\odot}$) & ($\mathrm{M_\odot}$)  & ($\mathrm{M_\odot}$)  &   & ($\mathrm{M_\odot}$)  &  & ($N_\mathrm{A} k_\mathrm{B}$)&$\mathrm{erg}$ & & ($\mathrm{M_\odot}$) & ($\mathrm{B}$) & ($\mathrm{km\,s^{-1}}$) & ($\mathrm{M_\odot}$) & ($\mathrm{M_\odot}$) & ($\mathrm{L_{p,0}}$)  & ($\mathrm{t_{p,0}}$)  & ($\log \mathrm{L_\odot}$)  & ($\mathrm{K}$) \\

        \midrule
        \endfirsthead
        \caption{continued.} \\
        \toprule
        $M_\mathrm{ini}$ & $\alpha_\mathrm{ov}$ & $M_\mathrm{He}$ & $M_\mathrm{CO}$  & $M_\mathrm{Fe}$  & $X_\mathrm{C}$  & $M_\mathrm{final}$  & $\xi_\mathrm{2.5}$ & $s_\mathrm{c}$ & $-E_\mathrm{bind}$ &Final fate? & $M_\mathrm{rm}$ & $E_\mathrm{expl}$ & $v_\mathrm{kick}$ & $M_\mathrm{Ni}$ & $M_\mathrm{ej}$  & $L_\mathrm{p}/\mathrm{L_{p,0}}$  & $t_\mathrm{p}/\mathrm{t_{p,0}}$  & $\log L_\mathrm{c}$  & $\log T_\mathrm{eff}$  \\
        
        ($\mathrm{M_\odot}$) & ($d_\mathrm{ov}/\mathrm{H_p}$) & ($\mathrm{M_\odot}$) & ($\mathrm{M_\odot}$)  & ($\mathrm{M_\odot}$)  &   & ($\mathrm{M_\odot}$)  &  & ($N_\mathrm{A} k_\mathrm{B}$) &$\mathrm{erg}$ & & ($\mathrm{M_\odot}$) & ($\mathrm{B}$) & ($\mathrm{km\,s^{-1}}$) & ($\mathrm{M_\odot}$) & ($\mathrm{M_\odot}$) & ($\mathrm{L_{p,0}}$)  & ($\mathrm{t_{p,0}}$)  & ($\log \mathrm{L_\odot}$)  & ($\mathrm{K}$) \\

        \midrule
        \endhead

        \toprule
        $M_\mathrm{ini}$ & $\alpha_\mathrm{ov}$ & $M_\mathrm{He}$ & $M_\mathrm{CO}$  & $M_\mathrm{Fe}$  & $X_\mathrm{C}$  & $M_\mathrm{final}$  & $\xi_\mathrm{2.5}$ & $s_\mathrm{c}$ & $-E_\mathrm{bind}$ &Final fate? & $M_\mathrm{rm}$ & $E_\mathrm{expl}$ & $v_\mathrm{kick}$ & $M_\mathrm{Ni}$ & $M_\mathrm{ej}$  & $L_\mathrm{p}/\mathrm{L_{p,0}}$  & $t_\mathrm{p}/\mathrm{t_{p,0}}$  & $\log L_\mathrm{c}$  & $\log T_\mathrm{eff}$  \\
        
        ($\mathrm{M_\odot}$) & ($d_\mathrm{ov}/\mathrm{H_p}$) & ($\mathrm{M_\odot}$) & ($\mathrm{M_\odot}$)  & ($\mathrm{M_\odot}$)  &   & ($\mathrm{M_\odot}$)  &  & ($N_\mathrm{A} k_\mathrm{B}$)&$\mathrm{erg}$ & & ($\mathrm{M_\odot}$) & ($\mathrm{B}$) & ($\mathrm{km\,s^{-1}}$) & ($\mathrm{M_\odot}$) & ($\mathrm{M_\odot}$) & ($\mathrm{L_{p,0}}$)  & ($\mathrm{t_{p,0}}$)  & ($\log \mathrm{L_\odot}$)  & ($\mathrm{K}$) \\
        
        \midrule
        \endhead

        \bottomrule
        \endfoot
        \input{data.tex}

        \end{longtable}
\end{landscape}
\twocolumn

%% file: data.tex
5.0 & 0.05 & 0.89 & 0.89 &-&- & 4.88 &-&-&-& WD &-&-&-&-&-&-&  & 4.42 & 3.5\\ 
5.0 & 0.1 & 0.91 & 0.91 &-&- & 4.86 &-&-&-& WD &-&-&-&-&-&-&-& 4.71 & 3.54\\ 
5.0 & 0.15 & 0.92 & 0.92 &-&- & 4.83 &-&-&-& WD &-&-&-&-&-&-&-& 4.83 & 3.55\\ 
5.0 & 0.2 & 0.93 & 0.92 &-&- & 4.76 &-&-&-& WD &-&-&-&-&-&-&-& 4.78 & 3.59\\ 
5.0 & 0.25 & 0.94 & 0.94 &-&- & 4.55 &-&-&-& WD &-&-&-&-&-&-&-& 5.07 & 3.57\\ 
5.0 & 0.3 & 0.95 & 0.95 &-&- & 4.29 &-&-&-& WD &-&-&-&-&-&-&-& 4.52 & 3.54\\ 
5.0 & 0.35 & 0.96 & 0.96 &-&- & 3.84 &-&-&-& WD &-&-&-&-&-&-&-& 5.7 & 3.67\\ 
5.0 & 0.4 & 0.98 & 0.98 &-&- & 4.6 &-&-&-& WD &-&-&-&-&-&-&-& 5.54 & 3.65\\ 
5.0 & 0.45 & 1.01 & 1.01 &-&- & 4.26 &-&-&-& WD &-&-&-&-&-&-&-& 5.95 & 3.69\\ 
5.0 & 0.5 & 1.05 & 1.05 &-&- & 4.04 &-&-&-& WD &-&-&-&-&-&-&-& 5.76 & 3.62\\ 

\midrule
6.0 & 0.05 & 0.95 & 0.95 &--&- & 5.82 &-&-&-& WD &-&-&-&-&-&-&-& 5.01 & 3.58\\ 
6.0 & 0.1 & 0.96 & 0.96 &-&- & 5.78 &-&-&-& WD &-&-&-&-&-&-&-& 5.12 & 3.59\\ 
6.0 & 0.15 & 0.98 & 0.98 &-&- & 5.62 &-&-&-& WD &-&-&-&-&-&-&-& 5.44 & 3.64\\ 
6.0 & 0.2 & 0.99 & 0.99 &-&- & 5.44 &-&-&-& WD &-&-&-&-&-&-&-& 5.66 & 3.66\\ 
6.0 & 0.25 & 1.01 & 1.01 &-&- & 5.27 &-&-&-& WD &-&-&-&-&-&-&-& 5.86 & 3.68\\ 
6.0 & 0.3 & 1.04 & 1.04 &-&- & 4.7 &-&-&-& WD &-&-&-&-&-&-&-& 6.18 & 3.71\\ 
6.0 & 0.35 & 1.09 & 1.09 &-&- & 3.59 &-&-&-& WD &-&-&-&-&-&-&-& 6.52 & 3.72\\ 
6.0 & 0.4 & 1.15 & 1.15 &-&- & 3.02 &-&-&-& WD &-&-&-&-&-&-&-& 6.54 & 3.71\\ 
6.0 & 0.45 & 1.22 & 1.22 &-&- & 4.34 &-&-&-& WD &-&-&-&-&-&-&-& 6.7 & 3.71\\ 
6.0 & 0.5 & 1.37 & 1.29 &-&- & 5.74 &-&-&-& WD &-&-&-&-&-&-&-& 5.39 & 3.72\\ 

\midrule
7.0 & 0.05 & 1.02 & 1.02 &-&- & 6.31 &-&-&-& WD &-&-&-&-&-&-&-& 5.83 & 3.68\\ 
7.0 & 0.1 & 1.04 & 1.04 &-&- & 5.48 &-&-&-& WD &-&-&-&-&-&-&-& 6.27 & 3.72\\ 
7.0 & 0.15 & 1.08 & 1.08 &-&- & 4.35 &-&-&-& WD &-&-&-&-&-&-&-& 6.55 & 3.72\\ 
7.0 & 0.2 & 1.13 & 1.13 &-&- & 3.82 &-&-&-& WD &-&-&-&-&-&-&-& 6.58 & 3.71\\ 
7.0 & 0.25 & 1.17 & 1.17 &-&- & 5.1 &-&-&-& WD &-&-&-&-&-&-&-& 6.5 & 3.73\\ 
7.0 & 0.3 & 1.23 & 1.22 &-&- & 6.74 &-&-&-& WD &-&-&-&-&-&-&-& 5.1 & 3.68\\ 
7.0 & 0.35 & 1.33 & 1.29 &-&- & 6.7 &-&-&-& WD &-&-&-&-&-&-&-& 5.43 & 3.73\\ 
7.0 & 0.4 & 1.36 & 1.36 &-&- & 6.66 &-&-&-& WD &-&-&-&-&-&-&-& 6.56 & 3.74\\ 
7.0 & 0.45 & 2.1 & 1.47 &-&- & 6.62 &-&-&-& Fe CCSN &-&-&-&-&-&-&-& 4.48 & 3.52 \\ 
7.0 & 0.5 & 2.48 & 1.54 &-&- & 6.57 &-&-&-& Fe CCSN &-&-&-&-&-&-&-& 4.57 & 3.51 \\ 

\midrule
8.0 & 0.05 & 1.14 & 1.14 &-&- & 4.66 &-&-&-& WD &-&-&-&-&-&-&-& 6.45 & 3.73 \\ 
8.0 & 0.1 & 1.18 & 1.18 &-&- & 6.53 &-&-&-& WD &-&-&-&-&-&-&-& 6.67 & 3.73 \\ 
8.0 & 0.15 & 1.23 & 1.23 &-&- & 6.93 &-&-&-& WD &-&-&-&-&-&-&-& 6.76 & 3.72 \\ 
8.0 & 0.2 & 1.31 & 1.28 &-&- & 7.69 &-&-&-& WD &-&-&-&-&-&-&-& 5.97 & 3.77 \\ 
8.0 & 0.25 & 1.34 & 1.34 &-&- & 7.65 &-&-&-& WD &-&-&-&-&-&-&-& 6.65 & 3.75 \\ 
8.0 & 0.3 & 2.03 & 1.42 &-&- & 7.59 &-&-&-& ECSN &-&-&-&-&-&-&-& 4.57 & 3.55 \\ 
8.0 & 0.35 & 2.5 & 1.5 &-&- & 7.53 &-&-&-& Fe CCSN &-&-&-&-&-&-&-& 4.53 & 3.52 \\ 
8.0 & 0.4 & 2.68 & 1.63 &-&- & 7.47 &-&-&-& Fe CCSN &-&-&-&-&-&-&-& 4.62 & 3.52 \\ 
8.0 & 0.45 & 2.84 & 1.75 &-&- & 7.39 &-&-&-& Fe CCSN &-&-&-&-&-&-&-& 4.67 & 3.51 \\ 
8.0 & 0.5 & 3.0 & 1.89 &-&- & 7.31 & 0.01 &-&-& Fe CCSN &-&-&-&-&-&-&-& 4.71 & 3.51 \\ 

\midrule
9.0 & 0.05 & 1.28 & 1.28 &-&- & 8.7 &-&-& -& WD &-&-&-&-&-&-&-& 5.74 & 3.71 \\ 
9.0 & 0.1 & 1.33 & 1.33 &-&- & 8.65 &-&-&-& WD &-&-&-&-&-&-&-& 7.82 & 3.7 \\ 
9.0 & 0.15 & 2.39 & 1.42 &-&- & 8.6 &-&-&-& ECSN &-&-&-&-&-&-&-& 4.56 & 3.56 \\ 
9.0 & 0.2 & 2.44 & 1.46 &-&- & 8.52 &-&-&-& Fe CCSN &-&-&-&-&-&-&-& 4.5 & 3.53 \\ 
9.0 & 0.25 & 2.64 & 1.54 &-&- & 8.46 &-&-&-& Fe CCSN &-&-&-&-&-&-&-& 4.57 & 3.53 \\ 
9.0 & 0.3 & 2.82 & 1.68 &-&- & 8.38 &-&-&-& Fe CCSN &-&-&-&-&-&-&-& 4.64 & 3.52 \\ 
9.0 & 0.35 & 2.98 & 1.82 &-&- & 8.29 & 0.01 &-&-& Fe CCSN &-&-&-&-&-&-&-& 4.68 & 3.52 \\ 
9.0 & 0.4 & 3.14 & 1.94 & 1.29 &- & 8.19 & 0.01 &-&-& Fe CCSN &-&-&-&-&-&-&-& 4.73 & 3.51 \\ 
9.0 & 0.45 & 3.31 & 1.97 & 1.49 & 0.35 & 8.08 & 0.02 & 0.78 & 50.73 & Fe CCSN & 1.28 & 1.0 & 580.75 & 0.03 & 6.66 & 1.65 & 0.88 & 4.78 & 3.51 \\  
9.0 & 0.5 & 3.48 & 2.13 & 1.66 & 0.35 & 7.96 & 0.03 & 0.8 & 50.54 & Fe CCSN & 1.34 & 0.83 & 497.76 & 0.03 & 6.47 & 1.51 & 0.9 & 4.83 & 3.51 \\  

\midrule
10.0 & 0.05 & 2.51 & 1.43 &-&- & 9.57 &-&-&-& Fe CCSN &-&-&-&-&-&-&-& 4.49 & 3.54 \\ 
10.0 & 0.1 & 2.63 & 1.5 &-&- & 9.49 &-&-&-& Fe CCSN &-&-&-&-&-&-&-& 4.55 & 3.54 \\ 
10.0 & 0.15 & 2.72 & 1.58 &-&- & 9.39 &-&-&-& Fe CCSN &-&-&-&-&-&-&-& 4.6 & 3.53 \\ 
10.0 & 0.2 & 2.9 & 1.7 &-&- & 9.3 &-&-&-& Fe CCSN &-&-&-&-&-&-&-& 4.65 & 3.53 \\ 
10.0 & 0.25 & 3.07 & 1.84 &-&- & 9.2 & 0.01 &-&-& Fe CCSN &-&-&-&-&-&-&-& 4.69 & 3.52 \\ 
10.0 & 0.3 & 3.24 & 1.85 & 1.51 & 0.34 & 9.08 & 0.01 & 0.83 & 50.75 & Fe CCSN & 1.35 & 0.8 & 381.26 & 0.03 & 7.58 & 1.22 & 0.96 & 4.75 & 3.52 \\  
10.0 & 0.35 & 3.43 & 2.01 & 1.52 & 0.34 & 8.95 & 0.02 & 0.82 & 50.76 & Fe CCSN & 1.35 & 0.78 & 384.21 & 0.03 & 7.45 & 1.27 & 0.97 & 4.8 & 3.51 \\  
10.0 & 0.4 & 3.61 & 2.18 & 1.59 & 0.34 & 8.81 & 0.03 & 0.82 & 50.66 & Fe CCSN & 1.36 & 0.73 & 396.1 & 0.03 & 7.3 & 1.27 & 0.98 & 4.85 & 3.51 \\  
10.0 & 0.45 & 3.81 & 2.36 & 1.57 & 0.34 & 8.65 & 0.05 & 0.83 & 50.78 & Fe CCSN & 1.38 & 0.71 & 367.75 & 0.03 & 7.12 & 1.32 & 0.98 & 4.89 & 3.51 \\  
10.0 & 0.5 & 4.0 & 2.55 & 1.54 & 0.34 & 8.49 & 0.07 & 0.85 & 50.9 & Fe CCSN & 1.38 & 0.74 & 363.22 & 0.03 & 6.94 & 1.43 & 0.97 & 4.92 & 3.5 \\

\midrule
11.0 & 0.05 & 2.77 & 1.61 &-&- & 10.35 &-&-&-& Fe CCSN &-&-&-&-&-&-&-& 4.61 & 3.54 \\ 
11.0 & 0.1 & 2.95 & 1.72 &-&- & 10.26 & 0.01 &-&-& Fe CCSN &-&-&-&-&-&-&-& 4.65 & 3.53 \\ 
11.0 & 0.15 & 3.13 & 1.88 &-&- & 10.15 & 0.01 &-&-& Fe CCSN &-&-&-&-&-&-&-& 4.7 & 3.53 \\ 
11.0 & 0.2 & 3.32 & 1.87 & 1.54 & 0.33 & 10.03 & 0.02 & 0.83 & 50.73 & Fe CCSN & 1.34 & 0.94 & 542.74 & 0.03 & 8.54 & 1.31 & 0.99 & 4.76 & 3.52 \\  
11.0 & 0.25 & 3.52 & 2.04 & 1.54 & 0.33 & 9.88 & 0.02 & 0.8 & 50.74 & Fe CCSN & 1.31 & 0.96 & 573.59 & 0.04 & 8.42 & 1.42 & 1.0 & 4.81 & 3.52 \\  
11.0 & 0.3 & 3.72 & 2.35 & 1.47 &- & 9.72 & 0.04 &-&-& Fe CCSN &-&-&-&-&-&-&-& 4.85 & 3.51 \\ 
11.0 & 0.35 & 3.92 & 2.41 & 1.56 & 0.33 & 9.55 & 0.06 & 0.82 & 50.82 & Fe CCSN & 1.34 & 0.87 & 443.23 & 0.04 & 8.05 & 1.46 & 1.01 & 4.9 & 3.51 \\  
11.0 & 0.4 & 4.14 & 2.6 & 1.56 & 0.32 & 9.36 & 0.09 & 0.87 & 50.96 & Fe CCSN & 1.41 & 0.86 & 428.83 & 0.04 & 7.78 & 1.52 & 1.01 & 4.93 & 3.51 \\  
11.0 & 0.45 & 4.36 & 2.8 & 1.51 & 0.32 & 9.15 & 0.09 & 0.83 & 50.96 & Fe CCSN & 1.36 & 0.81 & 406.58 & 0.04 & 7.65 & 1.51 & 1.01 & 4.97 & 3.51 \\  
11.0 & 0.5 & 4.56 & 3.01 & 1.53 & 0.32 & 8.94 & 0.09 & 0.84 & 50.97 & Fe CCSN & 1.37 & 0.79 & 400.18 & 0.04 & 7.41 & 1.54 & 1.01 & 5.01 & 3.52 \\  

\midrule
12.0 & 0.05 & 3.16 & 1.88 &-&- & 11.14 & 0.01 &-&-& Fe CCSN &-&-&-&-&-&-&-& 4.71 & 3.53 \\ 
12.0 & 0.1 & 3.35 & 1.87 & 1.61 & 0.31 & 11.01 & 0.02 & 0.77 & 50.44 & Fe CCSN & 1.3 & 0.85 & 512.99 & 0.03 & 9.57 & 1.12 & 1.06 & 4.76 & 3.53 \\  
12.0 & 0.15 & 3.56 & 2.18 & 1.62 &- & 10.86 & 0.02 &-&-& Fe CCSN &-&-&-&-&-&-&-& 4.81 & 3.52 \\ 
12.0 & 0.2 & 3.78 & 2.22 & 1.5 & 0.32 & 10.68 & 0.03 & 0.82 & 50.82 & Fe CCSN & 1.35 & 0.56 & 261.86 & 0.03 & 9.18 & 0.9 & 1.15 & 4.86 & 3.52 \\  
12.0 & 0.25 & 4.0 & 2.42 & 1.54 & 0.32 & 10.49 & 0.06 & 0.84 & 50.89 & Fe CCSN & 1.38 & 0.83 & 409.97 & 0.03 & 8.96 & 1.32 & 1.07 & 4.9 & 3.52 \\  
12.0 & 0.3 & 4.22 & 2.62 & 1.54 & 0.32 & 10.28 & 0.07 & 0.8 & 50.8 & Fe CCSN & 1.34 & 0.56 & 264.37 & 0.03 & 8.79 & 1.0 & 1.14 & 4.94 & 3.51 \\  
12.0 & 0.35 & 4.47 & 2.83 & 1.62 & 0.31 & 10.05 & 0.13 & 0.87 & 50.99 & Fe CCSN & 1.43 & 1.04 & 530.77 & 0.07 & 8.45 & 1.77 & 1.02 & 4.98 & 3.51 \\  
12.0 & 0.4 & 4.69 & 3.14 & 1.57 &- & 9.8 & 0.12 &-&-& Fe CCSN &-&-&-&-&-&-&-& 5.02 & 3.51 \\ 
12.0 & 0.45 & 4.89 & 3.27 & 1.63 & 0.31 & 9.55 & 0.16 & 0.91 & 51.09 & Fe CCSN & 1.46 & 0.99 & 498.53 & 0.07 & 7.91 & 1.76 & 1.0 & 5.02 & 3.52 \\  
12.0 & 0.5 & 5.11 & 3.49 & 1.63 & 0.31 & 9.29 & 0.18 & 0.92 & 51.17 & Fe CCSN & 1.49 & 1.24 & 626.55 & 0.09 & 7.61 & 2.16 & 0.94 & 5.07 & 3.53 \\

\midrule
13.0 & 0.05 & 3.56 & 2.03 & 1.54 & 0.28 & 11.86 & 0.02 & 0.82 & 50.69 & Fe CCSN & 1.34 & 0.76 & 374.32 & 0.03 & 10.38 & 1.02 & 1.14 & 4.8 & 3.53 \\  
13.0 & 0.1 & 3.78 & 2.19 & 1.54 & 0.3 & 11.69 & 0.04 & 0.8 & 50.78 & Fe CCSN & 1.33 & 0.9 & 541.59 & 0.04 & 10.22 & 1.23 & 1.11 & 4.85 & 3.52 \\  
13.0 & 0.15 & 4.01 & 2.39 & 1.55 & 0.31 & 11.49 & 0.06 & 0.82 & 50.83 & Fe CCSN & 1.34 & 0.9 & 456.64 & 0.04 & 9.99 & 1.31 & 1.11 & 4.89 & 3.52 \\  
13.0 & 0.2 & 4.25 & 2.59 & 1.58 & 0.31 & 11.27 & 0.09 & 0.86 & 50.98 & Fe CCSN & 1.44 & 0.84 & 457.65 & 0.05 & 9.65 & 1.31 & 1.12 & 4.93 & 3.52 \\  
13.0 & 0.25 & 4.49 & 2.81 & 1.57 & 0.3 & 11.02 & 0.13 & 0.8 & 51.0 & Fe CCSN & 1.34 & 1.48 & 951.86 & 0.07 & 9.53 & 2.2 & 1.02 & 4.97 & 3.51 \\  
13.0 & 0.3 & 4.74 & 3.04 & 1.6 & 0.3 & 10.76 & 0.14 & 0.9 & 51.08 & Fe CCSN & 1.46 & 1.09 & 539.07 & 0.07 & 9.11 & 1.77 & 1.05 & 5.01 & 3.52 \\  
13.0 & 0.35 & 4.97 & 3.27 & 1.53 & 0.3 & 10.48 & 0.15 & 0.8 & 51.08 & Fe CCSN & 1.33 & 1.21 & 757.14 & 0.07 & 9.0 & 1.97 & 1.03 & 5.04 & 3.52 \\  
13.0 & 0.4 & 5.21 & 3.5 & 1.71 & 0.3 & 10.18 & 0.19 & 0.92 & 51.12 & Fe CCSN & 1.52 & 1.33 & 683.4 & 0.11 & 8.47 & 2.19 & 0.99 & 5.07 & 3.53 \\  
13.0 & 0.45 & 5.45 & 3.74 & 1.69 & 0.3 & 9.88 & 0.23 & 0.95 & 51.22 & Fe CCSN & 1.56 & 1.41 & 703.29 & 0.11 & 8.11 & 2.36 & 0.96 & 5.11 & 3.54 \\  
13.0 & 0.5 & 5.69 & 3.98 & 1.77 & 0.3 & 9.56 & 0.24 & 0.96 & 51.21 & Fe CCSN & 1.6 & 1.54 & 770.03 & 0.12 & 7.75 & 2.57 & 0.92 & 5.13 & 3.54 \\

\midrule
15.0 & 0.05 & 4.4 & 2.66 & 1.6 & 0.28 & 13.16 & 0.1 & 0.87 & 50.97 & Fe CCSN & 1.45 & 0.82 & 421.39 & 0.05 & 11.54 & 1.16 & 1.22 & 4.95 & 3.52 \\  
15.0 & 0.1 & 4.68 & 2.9 & 1.53 & 0.29 & 12.86 & 0.08 & 0.79 & 50.88 & Fe CCSN & 1.32 & 0.76 & 375.8 & 0.04 & 11.4 & 1.14 & 1.24 & 4.99 & 3.52 \\  
15.0 & 0.15 & 4.95 & 3.13 & 1.61 & 0.29 & 12.56 & 0.16 & 0.9 & 51.1 & Fe CCSN & 1.46 & 1.1 & 552.5 & 0.07 & 10.92 & 1.64 & 1.15 & 5.03 & 3.52 \\  
15.0 & 0.2 & 5.23 & 3.38 & 1.6 & 0.29 & 12.21 & 0.17 & 0.8 & 51.08 & Fe CCSN & 1.35 & 1.6 & 1080.82 & 0.09 & 10.71 & 2.31 & 1.08 & 5.05 & 3.52 \\  
15.0 & 0.25 & 5.51 & 3.64 & 1.62 & 0.29 & 11.85 & 0.09 & 0.84 & 50.9 & Fe CCSN & 1.38 & 0.52 & 266.67 & 0.03 & 10.31 & 0.92 & 1.28 & 5.09 & 3.53 \\  
15.0 & 0.3 & 5.79 & 3.91 & 1.62 & 0.29 & 11.46 & 0.11 & 0.83 & 50.98 & Fe CCSN & 1.4 & 0.6 & 292.3 & 0.04 & 9.9 & 1.07 & 1.23 & 5.13 & 3.54 \\  
15.0 & 0.35 & 6.07 & 4.18 & 1.6 & 0.29 & 11.06 & 0.15 & 0.9 & 51.16 & Fe CCSN & 1.45 & 0.72 & 356.37 & 0.05 & 9.43 & 1.28 & 1.16 & 5.16 & 3.54 \\  
15.0 & 0.4 & 6.34 & 4.46 & 1.74 & 0.29 & 10.64 & 0.15 & 0.91 & 51.09 & Fe CCSN & 1.48 & 0.67 & 342.63 & 0.07 & 8.99 & 1.21 & 1.14 & 5.19 & 3.56 \\  
15.0 & 0.45 & 6.62 & 4.74 & 1.67 & 0.29 & 10.22 & 0.21 & 0.96 & 51.3 & Fe CCSN & 1.55 & 0.96 & 465.02 & 0.09 & 8.47 & 1.66 & 1.04 & 5.22 & 3.57 \\  
15.0 & 0.5 & 6.89 & 5.01 & 1.72 & 0.29 & 9.8 & 0.22 & 0.97 & 51.31 & Fe CCSN & 1.57 & 0.94 & 446.07 & 0.09 & 8.03 & 1.62 & 1.01 & 5.24 & 3.58 \\

\midrule
17.0 & 0.05 & 5.28 & 3.37 & 1.75 & 0.27 & 14.22 & 0.12 & 0.84 & 50.88 & Fe CCSN & 1.41 & 0.87 & 512.17 & 0.06 & 12.64 & 1.27 & 1.3 & 5.06 & 3.52 \\  
17.0 & 0.1 & 5.61 & 3.65 & 1.55 & 0.27 & 13.78 & 0.12 & 0.87 & 51.07 & Fe CCSN & 1.41 & 0.63 & 306.79 & 0.04 & 12.2 & 1.02 & 1.35 & 5.09 & 3.52 \\  
17.0 & 0.15 & 5.94 & 3.94 & 1.62 & 0.28 & 13.3 & 0.16 & 0.92 & 51.17 & Fe CCSN & 1.47 & 1.19 & 628.64 & 0.06 & 11.65 & 1.77 & 1.19 & 5.14 & 3.53 \\  
17.0 & 0.2 & 6.26 & 4.24 & 1.69 & 0.28 & 12.81 & 0.23 & 0.97 & 51.28 & Fe CCSN & 1.57 & 1.13 & 544.46 & 0.11 & 11.03 & 1.75 & 1.17 & 5.18 & 3.54 \\  
17.0 & 0.25 & 6.58 & 4.54 & 1.73 & 0.28 & 12.31 & 0.24 & 0.99 & 51.3 & Fe CCSN & 1.6 & 1.12 & 535.36 & 0.11 & 10.5 & 1.77 & 1.14 & 5.21 & 3.55 \\  
17.0 & 0.3 & 6.9 & 4.85 & 1.68 & 0.27 & 11.76 & 0.17 & 0.92 & 51.21 & Fe CCSN & 1.48 & 0.69 & 346.91 & 0.06 & 10.1 & 1.19 & 1.21 & 5.24 & 3.56 \\  
17.0 & 0.35 & 7.22 & 5.16 & 1.83 & 0.27 & 11.24 & 0.31 & 1.02 & 51.38 & Fe CCSN & 1.65 & 1.99 & 1008.01 & 0.15 & 9.36 & 2.94 & 0.97 & 5.27 & 3.58 \\  
17.0 & 0.4 & 7.53 & 5.48 & 1.71 & 0.27 & 10.7 & 0.28 & 0.98 & 51.42 & Fe CCSN & 1.59 & 1.63 & 871.13 & 0.12 & 8.9 & 2.47 & 0.97 & 5.3 & 3.6 \\  
17.0 & 0.45 & 7.85 & 5.79 & 1.61 & 0.27 & 10.23 & 0.2 & 0.9 & 51.36 & Fe CCSN & 1.44 & 1.1 & 658.83 & 0.08 & 8.61 & 1.72 & 1.01 & 5.32 & 3.62 \\  
17.0 & 0.5 & 8.15 & 6.1 & 1.61 & 0.27 & 9.84 & 0.19 & 0.91 & 51.38 & Fe CCSN & 1.46 & 0.82 & 468.84 & 0.08 & 8.2 & 1.29 & 1.02 & 5.34 & 3.64 \\

\midrule
18.0 & 0.05 & 5.75 & 3.76 & 1.82 & 0.26 & 14.66 & 0.15 & 0.85 & 50.97 & Fe CCSN & 1.45 & 0.97 & 576.48 & 0.08 & 13.04 & 1.4 & 1.3 & 5.13 & 3.53 \\  
18.0 & 0.1 & 6.09 & 4.06 & 1.68 & 0.26 & 14.11 & 0.2 & 0.94 & 51.24 & Fe CCSN & 1.59 & 1.04 & 535.8 & 0.1 & 12.3 & 1.54 & 1.25 & 5.16 & 3.54 \\  
18.0 & 0.15 & 6.45 & 4.36 & 1.69 & 0.27 & 13.56 & 0.22 & 0.96 & 51.28 & Fe CCSN & 1.57 & 1.42 & 713.26 & 0.1 & 11.78 & 2.06 & 1.16 & 5.2 & 3.54 \\  
18.0 & 0.2 & 6.79 & 4.68 & 1.7 & 0.27 & 12.99 & 0.23 & 0.96 & 51.31 & Fe CCSN & 1.55 & 1.42 & 734.75 & 0.1 & 11.23 & 2.11 & 1.14 & 5.22 & 3.55 \\  
18.0 & 0.25 & 7.13 & 5.01 & 1.6 & 0.27 & 12.4 & 0.22 & 0.9 & 51.34 & Fe CCSN & 1.43 & 1.64 & 1034.02 & 0.1 & 10.8 & 2.39 & 1.08 & 5.26 & 3.56 \\  
18.0 & 0.3 & 7.48 & 5.34 & 1.73 & 0.27 & 11.77 & 0.24 & 0.94 & 51.34 & Fe CCSN & 1.51 & 1.44 & 812.89 & 0.1 & 10.07 & 2.18 & 1.06 & 5.29 & 3.58 \\  
18.0 & 0.35 & 7.82 & 5.67 & 1.65 & 0.27 & 11.18 & 0.17 & 0.88 & 51.26 & Fe CCSN & 1.42 & 0.93 & 536.4 & 0.07 & 9.6 & 1.49 & 1.11 & 5.32 & 3.6 \\  
18.0 & 0.4 & 8.15 & 6.0 & 1.63 & 0.26 & 10.67 & 0.22 & 0.93 & 51.41 & Fe CCSN & 1.48 & 0.96 & 548.81 & 0.09 & 9.0 & 1.5 & 1.05 & 5.34 & 3.62 \\  
18.0 & 0.45 & 8.48 & 6.34 & 1.61 & 0.26 & 10.24 & 0.15 & 0.85 & 51.26 & Fe CCSN & 1.38 & 0.75 & 460.77 & 0.06 & 8.7 & 1.16 & 1.06 & 5.38 & 3.65 \\  
18.0 & 0.5 & 8.8 & 6.67 & 1.72 & 0.26 & 9.94 & 0.26 & 0.98 & 51.46 & Fe CCSN & 1.58 & 1.27 & 683.09 & 0.11 & 8.15 & 1.71 & 0.92 & 5.39 & 3.68 \\

\midrule

19.0 & 0.05 & 6.26 & 4.16 & 1.71 & 0.25 & 14.92 & 0.17 & 0.93 &-& Fe CCSN & 1.56 & 0.91 & 480.95 & 0.09 & 13.16 & 1.34 & 1.33 & 5.18 & 3.54 \\ 
19.0 & 0.1 & 6.6 & 4.47 & 1.69 & 0.26 & 14.35 & 0.21 & 0.94 & 51.27 & Fe CCSN & 1.57 & 1.17 & 635.6 & 0.11 & 12.57 & 1.71 & 1.24 & 5.2 & 3.54 \\  
19.0 & 0.15 & 6.97 & 4.81 & 1.76 & 0.26 & 13.69 & 0.26 & 0.99 & 51.33 & Fe CCSN & 1.7 & 1.05 & 499.13 & 0.11 & 11.76 & 1.62 & 1.23 & 5.25 & 3.55 \\  
19.0 & 0.2 & 7.33 & 5.15 & 1.73 & 0.26 & 13.05 & 0.26 & 0.99 & 51.37 & Fe CCSN & 1.62 & 1.62 & 811.64 & 0.12 & 11.2 & 2.33 & 1.11 & 5.28 & 3.57 \\  
19.0 & 0.25 & 7.7 & 5.49 & 1.72 & 0.26 & 12.38 & 0.2 & 0.91 & 51.29 & Fe CCSN & 1.48 & 1.02 & 576.36 & 0.08 & 10.72 & 1.61 & 1.17 & 5.31 & 3.58 \\  
19.0 & 0.3 & 8.06 & 5.84 & 2.02 & 0.26 & 11.74 & 0.48 & 1.08 & 51.52 & Fe CCSN & 1.87 & 3.46 & 1673.73 & 0.24 & 9.57 & 4.54 & 0.89 & 5.33 & 3.6 \\  
19.0 & 0.35 & 8.42 & 6.19 & 1.66 & 0.26 & 11.14 & 0.14 & 0.88 & 51.25 & Fe CCSN & 1.42 & 0.83 & 472.5 & 0.06 & 9.55 & 1.31 & 1.11 & 5.37 & 3.62 \\  
19.0 & 0.4 & 8.77 & 6.55 & 1.72 & 0.26 & 10.68 & 0.21 & 0.93 & 51.37 & Fe CCSN & 1.49 & 0.99 & 563.71 & 0.09 & 9.01 & 1.46 & 1.03 & 5.37 & 3.65 \\  
19.0 & 0.45 & 9.1 & 6.89 & 2.19 & 0.26 & 10.36 & 0.68 & 1.17 & 51.68 & Fe CCSN & 10.36 &-&-&-&- &-&-& 5.41 & 3.68 \\  
19.0 & 0.5 & 9.43 & 7.25 & 2.1 & 0.25 & 10.17 & 0.65 & 1.13 & 51.67 & Fe CCSN & 10.17 &-&-&-&- &-&-& 5.44 & 3.73 \\

\midrule
20.0 & 0.05 & 6.79 & 4.61 & 1.66 & 0.24 & 14.91 & 0.16 & 0.88 & 51.17 & Fe CCSN & 1.42 & 0.94 & 528.38 & 0.07 & 13.32 & 1.39 & 1.33 & 5.23 & 3.55 \\  
20.0 & 0.1 & 7.14 & 4.93 & 1.72 & 0.25 & 14.43 & 0.25 & 0.91 & 51.3 & Fe CCSN & 1.54 & 2.18 & 1380.18 & 0.18 & 12.68 & 2.86 & 1.13 & 5.25 & 3.55 \\  
20.0 & 0.15 & 7.51 & 5.26 & 1.56 & 0.25 & 13.74 & 0.16 & 0.88 & 51.28 & Fe CCSN & 1.4 & 0.86 & 490.66 & 0.06 & 12.17 & 1.34 & 1.29 & 5.29 & 3.57 \\  
20.0 & 0.2 & 7.9 & 5.62 & 2.0 & 0.25 & 13.0 & 0.48 & 1.08 & 51.51 & Fe CCSN & 1.87 & 3.4 & 1633.1 & 0.26 & 10.84 & 4.38 & 0.96 & 5.31 & 3.58 \\  
20.0 & 0.25 & 8.28 & 5.98 & 1.58 & 0.25 & 12.31 & 0.13 & 0.83 & 51.24 & Fe CCSN & 1.34 & 1.19 & 823.51 & 0.07 & 10.82 & 1.75 & 1.13 & 5.35 & 3.6 \\  
20.0 & 0.3 & 8.66 & 6.35 & 1.78 & 0.25 & 11.66 & 0.35 & 1.0 & 51.47 & Fe CCSN & 1.76 & 1.73 & 948.76 & 0.16 & 9.64 & 2.43 & 0.99 & 5.38 & 3.62 \\  
20.0 & 0.35 & 9.02 & 6.72 & 2.14 & 0.25 & 11.15 & 0.69 & 1.17 & 51.66 & Fe CCSN & 11.15 &-&-&-&- &-&-& 5.4 & 3.65 \\  
20.0 & 0.4 & 9.38 & 7.09 & 2.02 & 0.25 & 10.79 & 0.52 & 1.13 & 51.63 & Fe CCSN & 10.79 &-&-&-&- &-&-& 5.43 & 3.68 \\  
20.0 & 0.45 & 9.7 & 7.44 & 2.08 & 0.25 & 10.62 & 0.61 & 1.11 & 51.67 & Fe CCSN & 10.62 &-&-&-&- &-&-& 5.46 & 3.72 \\  
20.0 & 0.5 & 9.98 & 7.75 & 2.03 & 0.25 & 10.58 & 0.5 & 1.08 & 51.65 & Fe CCSN & 1.93 & 3.11 & 1515.54 & 0.24 & 8.33 & 3.1 & 0.77 & 5.49 & 3.76 \\  

\midrule
21.0 & 0.05 & 7.29 & 5.04 & 1.61 & 0.24 & 14.98 & 0.11 & 0.84 & 51.08 & Fe CCSN & 1.38 & 0.43 & 241.4 & 0.03 & 13.45 & 0.72 & 1.53 & 5.28 & 3.56 \\  
21.0 & 0.1 & 7.69 & 5.38 & 1.82 & 0.24 & 14.39 & 0.34 & 1.04 & 51.45 & Fe CCSN & 1.72 & 1.7 & 791.44 & 0.16 & 12.42 & 2.38 & 1.16 & 5.31 & 3.57 \\  
21.0 & 0.15 & 8.07 & 5.74 & 1.86 & 0.24 & 13.66 & 0.14 & 0.86 & 51.11 & Fe CCSN & 1.41 & 1.1 & 719.49 & 0.08 & 12.08 & 1.63 & 1.23 & 5.33 & 3.58 \\  
21.0 & 0.2 & 8.48 & 6.11 & 1.63 & 0.25 & 12.88 & 0.18 & 0.92 & 51.35 & Fe CCSN & 1.48 & 0.92 & 510.59 & 0.08 & 11.22 & 1.41 & 1.21 & 5.35 & 3.6 \\  
21.0 & 0.25 & 8.86 & 6.49 & 2.02 & 0.25 & 12.2 & 0.52 & 1.12 & 51.58 & Fe CCSN & 1.96 & 3.9 & 1931.84 & 0.26 & 9.93 & 4.83 & 0.89 & 5.39 & 3.62 \\  
21.0 & 0.3 & 9.25 & 6.88 & 2.1 & 0.24 & 11.65 & 0.56 & 1.14 & 51.63 & Fe CCSN & 11.65 &-&-&-&- &-&-& 5.41 & 3.64 \\  
21.0 & 0.35 & 9.58 & 7.22 & 2.11 & 0.25 & 11.39 & 0.61 & 1.12 & 51.64 & Fe CCSN & 11.39 &-&-&-&- &-&-& 5.45 & 3.67 \\  
21.0 & 0.4 & 9.88 & 7.55 & 2.08 & 0.25 & 11.22 & 0.54 & 1.08 & 51.64 & Fe CCSN & 1.98 & 3.18 & 1540.57 & 0.25 & 8.91 & 3.64 & 0.83 & 5.48 & 3.7 \\  
21.0 & 0.45 & 10.24 & 7.94 & 1.61 & 0.24 & 10.98 & 0.18 & 0.92 & 51.48 & Fe CCSN & 1.44 & 0.4 & 261.56 & 0.05 & 9.36 & 0.55 & 1.17 & 5.51 & 3.76 \\  
21.0 & 0.5 & 10.6 & 8.33 & 1.74 & 0.24 & 10.94 & 0.16 & 0.92 & 51.42 & Fe CCSN & 1.48 & 0.25 & 191.76 & 0.06 & 9.27 & 0.23 & 1.11 & 5.53 & 3.92 \\  

\midrule
22.0 & 0.05 & 7.9 & 5.52 & 1.7 & 0.23 & 14.74 & 0.19 & 0.94 & 51.31 & Fe CCSN & 1.57 & 1.06 & 620.38 & 0.1 & 12.97 & 1.56 & 1.29 & 5.32 & 3.57 \\  
22.0 & 0.1 & 8.26 & 5.87 & 1.64 & 0.24 & 14.11 & 0.2 & 0.93 & 51.36 & Fe CCSN & 1.5 & 1.01 & 551.39 & 0.08 & 12.42 & 1.51 & 1.26 & 5.34 & 3.58 \\  
22.0 & 0.15 & 8.64 & 6.23 & 1.94 & 0.24 & 13.43 & 0.33 & 1.02 & 51.43 & Fe CCSN & 1.67 & 1.8 & 935.29 & 0.15 & 11.53 & 2.48 & 1.1 & 5.36 & 3.59 \\  
22.0 & 0.2 & 9.05 & 6.61 & 2.23 & 0.24 & 12.75 & 0.67 & 1.16 & 51.65 & Fe CCSN & 12.75 &-&-&-&- &-&-& 5.4 & 3.62 \\  
22.0 & 0.25 & 9.44 & 7.0 & 2.08 & 0.24 & 12.18 & 0.63 & 1.12 & 51.65 & Fe CCSN & 12.18 &-&-&-&- &-&-& 5.43 & 3.64 \\  
22.0 & 0.3 & 9.75 & 7.32 & 2.08 & 0.24 & 12.05 & 0.57 & 1.1 & 51.64 & Fe CCSN & 12.05 &-&-&-&- &-&-& 5.46 & 3.66 \\  
22.0 & 0.35 & 10.1 & 7.7 & 2.01 & 0.24 & 11.7 & 0.46 & 1.07 & 51.65 & Fe CCSN & 1.86 & 3.76 & 1929.6 & 0.26 & 9.55 & 4.15 & 0.84 & 5.49 & 3.69 \\  
22.0 & 0.4 & 10.48 & 8.11 & 1.67 & 0.24 & 11.41 & 0.19 & 0.92 & 51.51 & Fe CCSN & 1.5 & 0.39 & 246.3 & 0.07 & 9.72 & 0.55 & 1.21 & 5.52 & 3.74 \\  
22.0 & 0.45 & 10.86 & 8.53 & 1.74 & 0.24 & 11.3 & 0.23 & 0.93 & 51.55 & Fe CCSN & 1.52 & 0.89 & 545.3 & 0.12 & 9.59 & 0.77 & 0.95 & 5.56 & 3.88 \\  
22.0 & 0.5 & 11.42 & 8.93 &-& - & 11.48 & 0.03 & - &0& Fe CCSN &-&-&-&-&-&-&-&  5.51 & 3.98 \\ 
\midrule
23.0 & 0.05 & 8.4 & 5.99 & 2.12 & 0.22 & 14.47 & 0.6 & 1.13 & 51.56 & Fe CCSN & 14.47 &-&-&-&- &-&-& 5.36 & 3.58 \\  
23.0 & 0.1 & 8.81 & 6.35 & 2.2 & 0.23 & 13.88 & 0.68 & 1.17 & 51.64 & Fe CCSN & 13.88 &-&-&-&- &-&-& 5.39 & 3.6 \\  
23.0 & 0.15 & 9.2 & 6.7 & 2.1 & 0.23 & 13.41 & 0.63 & 1.12 & 51.64 & Fe CCSN & 13.41 &-&-&-&- &-&-& 5.42 & 3.61 \\  
23.0 & 0.2 & 9.6 & 7.1 & 2.03 & 0.23 & 12.79 & 0.57 & 1.1 & 51.64 & Fe CCSN & 12.79 &-&-&-&- &-&-& 5.44 & 3.63 \\  
23.0 & 0.25 & 9.91 & 7.41 & 2.06 & 0.24 & 12.66 & 0.51 & 1.1 & 51.62 & Fe CCSN & 1.94 & 3.02 & 1457.9 & 0.24 & 10.41 & 3.7 & 0.94 & 5.47 & 3.65 \\  
23.0 & 0.3 & 10.3 & 7.84 & 1.72 & 0.24 & 12.19 & 0.29 & 0.98 & 51.6 & Fe CCSN & 1.56 & 1.34 & 803.05 & 0.13 & 10.42 & 1.72 & 1.05 & 5.51 & 3.68 \\  
23.0 & 0.35 & 10.69 & 8.26 & 1.61 & 0.24 & 11.87 & 0.17 & 0.92 & 51.5 & Fe CCSN & 1.48 & 0.35 & 229.64 & 0.06 & 10.2 & 0.51 & 1.27 & 5.54 & 3.73 \\  
23.0 & 0.4 & 11.1 & 8.69 & 1.7 & 0.24 & 11.7 & 0.24 & 0.94 & 51.59 & Fe CCSN & 1.55 & 1.0 & 596.17 & 0.13 & 9.95 & 0.95 & 0.99 & 5.58 & 3.83 \\  
23.0 & 0.45 & 11.78 & 9.12 & 1.59 & 0.23 & 11.78 & 0.14 & 0.84 & 51.44 & Fe CCSN & 1.36 & 0.61 & 420.24 & 0.06 & 10.27 & 0.31 & 0.91 & 5.59 & 4.06 \\  
23.0 & 0.5 & 12.12 & 9.54 &-&- & 12.13 & 0.03 &-&-& Fe CCSB &-&-&-&-&-&-&-& 5.54 & 4.41 \\ 

\midrule
24.0 & 0.05 & 9.07 & 6.46 & 2.06 & 0.21 & 14.13 & 0.57 & 1.09 & 51.6 & Fe CCSN & 14.13 &-&-&-&- &-&-& 5.39 & 3.6 \\  
24.0 & 0.1 & 9.36 & 6.81 & 2.06 & 0.22 & 13.78 & 0.57 & 1.1 & 51.62 & Fe CCSN & 13.78 &-&-&-&- &-&-& 5.42 & 3.62 \\  
24.0 & 0.15 & 9.75 & 7.22 & 2.02 & 0.22 & 13.28 & 0.5 & 1.08 & 51.63 & Fe CCSN & 13.28 &-&-&-&- &-&-& 5.44 & 3.63 \\  
24.0 & 0.2 & 10.04 & 7.52 & 1.86 & 0.23 & 13.21 & 0.38 & 1.04 & 51.61 & Fe CCSN & 1.75 & 2.46 & 1256.01 & 0.19 & 11.2 & 3.04 & 1.01 & 5.49 & 3.65 \\  
24.0 & 0.25 & 10.48 & 7.94 & 1.62 & 0.23 & 12.71 & 0.17 & 0.91 & 51.48 & Fe CCSN & 1.47 & 0.33 & 216.51 & 0.05 & 11.05 & 0.54 & 1.38 & 5.52 & 3.68 \\  
24.0 & 0.3 & 10.86 & 8.38 & 1.68 & 0.23 & 12.35 & 0.24 & 0.95 & 51.58 & Fe CCSN & 1.55 & 1.02 & 591.5 & 0.12 & 10.6 & 1.27 & 1.09 & 5.55 & 3.72 \\  
24.0 & 0.35 & 11.29 & 8.83 & 1.57 & 0.23 & 12.13 & 0.16 & 0.84 & 51.48 & Fe CCSN & 1.34 & 0.99 & 727.71 & 0.07 & 10.63 & 0.99 & 1.04 & 5.59 & 3.79 \\  
24.0 & 0.4 & 11.72 & 9.28 & 1.69 & 0.23 & 12.15 & 0.19 & 0.94 & 51.53 & Fe CCSN & 1.55 & 0.84 & 569.89 & 0.1 & 10.4 & 0.59 & 0.96 & 5.6 & 3.93 \\  
24.0 & 0.45 & 12.38 & 9.72 & - & - & 12.39 & 0.03 & - &-& Fe CCSN &-&-&-&-&-&-&-& 5.55 & 4.41 \\ 
24.0 & 0.5 & 12.69 & 10.16 & 1.89 & 0.23 & 12.69 & 0.22 & 0.97 & 51.54 & Fe CCSN & 1.62 & 1.06 & 625.46 & 0.13 & 10.86 & 0.25 & 0.73 & 5.65 & 4.3 \\  

\midrule
25.0 & 0.05 & 9.57 & 6.9 & 1.94 & 0.21 & 14.14 & 0.46 & 1.08 & 51.61 & Fe CCSN & 1.86 & 3.04 & 1475.49 & 0.23 & 11.99 & 3.74 & 1.03 & 5.43 & 3.61 \\  
25.0 & 0.1 & 9.92 & 7.27 & 1.87 & 0.22 & 13.92 & 0.4 & 1.06 & 51.62 & Fe CCSN & 1.77 & 2.68 & 1345.59 & 0.21 & 11.88 & 3.32 & 1.04 & 5.46 & 3.63 \\  
25.0 & 0.15 & 10.2 & 7.64 & 1.7 & 0.22 & 13.77 & 0.24 & 0.98 & 51.54 & Fe CCSN & 1.6 & 0.65 & 351.99 & 0.11 & 11.96 & 0.99 & 1.31 & 5.5 & 3.64 \\  
25.0 & 0.2 & 10.6 & 8.02 & 1.63 & 0.23 & 13.32 & 0.18 & 0.92 & 51.48 & Fe CCSN & 1.5 & 0.31 & 205.36 & 0.06 & 11.64 & 0.51 & 1.44 & 5.53 & 3.67 \\  
25.0 & 0.25 & 11.05 & 8.49 & 1.64 & 0.22 & 12.88 & 0.17 & 0.87 & 51.48 & Fe CCSN & 1.4 & 0.32 & 239.31 & 0.06 & 11.31 & 0.48 & 1.39 & 5.56 & 3.7 \\  
25.0 & 0.3 & 11.45 & 8.96 & 1.69 & 0.22 & 12.6 & 0.21 & 0.95 & 51.54 & Fe CCSN & 1.58 & 0.73 & 491.27 & 0.1 & 10.82 & 0.83 & 1.13 & 5.6 & 3.77 \\  
25.0 & 0.35 & 11.9 & 9.41 & 1.83 & 0.22 & 12.56 & 0.19 & 0.94 & 51.49 & Fe CCSN & 1.56 & 0.89 & 539.95 & 0.11 & 10.8 & 0.65 & 0.98 & 5.64 & 3.91 \\  
25.0 & 0.4 & 12.73 & 9.87 & 1.58 & 0.22 & 12.73 & 0.16 & 0.84 & 51.52 & Fe CCSN & 1.36 & 0.78 & 581.89 & 0.07 & 11.22 & 0.4 & 0.94 & 5.64 & 4.05 \\  
25.0 & 0.45 & 13.07 & 10.33 & - &-& 13.08 & 0.03 &-&-& Fe CCSN &-&-&-&-&-&-&-& 5.59 & 4.41 \\ 
25.0 & 0.5 & 13.3 & 10.78 & 1.69 & 0.22 & 13.3 & 0.18 & 0.9 & 51.56 & Fe CCSN & 1.44 & 0.45 & 341.35 & 0.07 & 11.68 & 0.09 & 0.82 & 5.69 & 4.4 \\

\midrule
27.0 & 0.05 & 10.64 & 7.9 & 1.71 & 0.2 & 14.39 & 0.24 & 0.98 & 51.56 & Fe CCSN & 1.6 & 0.62 & 341.13 & 0.11 & 12.57 & 0.93 & 1.36 & 5.51 & 3.65 \\  
27.0 & 0.1 & 10.94 & 8.19 & 1.63 & 0.21 & 14.5 & 0.21 & 0.93 & 51.55 & Fe CCSN & 1.49 & 0.52 & 326.15 & 0.08 & 12.82 & 0.79 & 1.4 & 5.55 & 3.66 \\  
27.0 & 0.15 & 11.23 & 8.61 & 1.77 & 0.22 & 14.3 & 0.18 & 0.92 & 51.45 & Fe CCSN & 1.51 & 0.83 & 503.23 & 0.09 & 12.6 & 1.07 & 1.26 & 5.57 & 3.69 \\  
27.0 & 0.2 & 11.75 & 9.1 & 1.8 & 0.22 & 13.89 & 0.19 & 0.94 & 51.48 & Fe CCSN & 1.53 & 0.83 & 507.67 & 0.1 & 12.16 & 1.03 & 1.22 & 5.6 & 3.72 \\  
27.0 & 0.25 & 12.18 & 9.6 & 1.59 & 0.22 & 13.58 & 0.16 & 0.86 & 51.52 & Fe CCSN & 1.39 & 0.63 & 454.75 & 0.07 & 12.03 & 0.67 & 1.21 & 5.65 & 3.79 \\  
27.0 & 0.3 & 12.65 & 10.08 & 1.73 & 0.21 & 13.46 & 0.28 & 0.95 & 51.65 & Fe CCSN & 1.7 & 1.08 & 694.61 & 0.14 & 11.51 & 0.76 & 0.99 & 5.65 & 3.92 \\  
27.0 & 0.35 & 13.58 & 10.57 & 1.69 & 0.21 & 13.58 & 0.24 & 0.98 & 51.63 & Fe CCSN & 1.57 & 0.93 & 571.85 & 0.11 & 11.81 & 0.39 & 0.9 & 5.67 & 4.11 \\  
27.0 & 0.4 & 13.96 & 11.06 & - & - & 13.97 & 0.02 &-&-& Fe CCSN &-&-&-&-&-&-&-& 5.59 & 4.62 \\ 
27.0 & 0.45 & 14.15 & 11.54 & 1.8 & 0.21 & 14.15 & 0.25 & 0.96 & 51.65 & Fe CCSN & 1.54 & 0.84 & 550.79 & 0.1 & 12.4 & 0.13 & 0.73 & 5.72 & 4.46 \\  
27.0 & 0.5 & 14.39 & 12.02 & 1.89 & 0.21 & 14.39 & 0.39 & 1.05 & 51.74 & Fe CCSN & 1.76 & 1.92 & 1049.53 & 0.19 & 12.37 & 0.17 & 0.57 & 5.74 & 4.59 \\  

\midrule
30.0 & 0.05 & 12.06 & 9.24 & 1.85 & 0.19 & 15.19 & 0.24 & 0.96 & 51.55 & Fe CCSN & 1.58 & 1.02 & 613.74 & 0.11 & 13.4 & 1.25 & 1.26 & 5.62 & 3.7 \\  
30.0 & 0.1 & 12.48 & 9.68 & 1.75 & 0.2 & 15.05 & 0.23 & 0.98 & 51.59 & Fe CCSN & 1.58 & 0.86 & 531.2 & 0.11 & 13.27 & 1.03 & 1.27 & 5.65 & 3.73 \\  
30.0 & 0.15 & 15.06 & 10.14 & 1.82 & 0.16 & 15.06 & 0.36 & 1.06 & 51.73 & Fe CCSN & 1.69 & 1.92 & 1088.01 & 0.17 & 13.14 & 1.08 & 0.95 & 5.72 & 3.95 \\  
30.0 & 0.2 & 13.38 & 10.73 & 1.64 & 0.2 & 14.6 & 0.26 & 0.93 & 51.66 & Fe CCSN & 1.47 & 1.63 & 1155.61 & 0.12 & 12.95 & 1.05 & 0.99 & 5.68 & 3.92 \\  
30.0 & 0.25 & 14.66 & 11.25 & 1.9 & 0.2 & 14.66 & 0.3 & 0.98 & 51.67 & Fe CCSN & 1.62 & 1.22 & 752.34 & 0.13 & 12.82 & 0.65 & 0.97 & 5.71 & 4.01 \\  
30.0 & 0.3 & 14.92 & 11.8 & 1.84 & 0.2 & 14.92 & 0.39 & 1.07 & 51.76 & Fe CCSN & 1.76 & 1.92 & 1051.35 & 0.18 & 12.9 & 0.51 & 0.78 & 5.74 & 4.21 \\  
30.0 & 0.35 & 15.24 & 12.32 & 1.88 & 0.2 & 15.24 & 0.3 & 0.99 & 51.72 & Fe CCSN & 1.61 & 1.1 & 706.77 & 0.13 & 13.41 & 0.19 & 0.76 & 5.76 & 4.39 \\  
30.0 & 0.4 & 15.52 & 12.86 & 1.92 & 0.2 & 15.52 & 0.45 & 1.1 & 51.82 & Fe CCSN & 1.83 & 2.34 & 1260.2 & 0.21 & 13.41 & 0.23 & 0.61 & 5.77 & 4.53 \\  
30.0 & 0.45 & 15.77 & 13.37 & 2.09 & 0.2 & 15.77 & 0.44 & 1.09 & 51.81 & Fe CCSN & 1.8 & 2.22 & 1225.6 & 0.2 & 13.69 & 0.08 & 0.48 & 5.8 & 4.89 \\  
30.0 & 0.5 & 15.96 & 13.86 & 2.01 & 0.2 & 15.96 & 0.59 & 1.14 & 51.88 & Fe CCSN & 15.96 &-&-&-&- &-&-& 5.8 & 5.33 \\

\midrule
35.0 & 0.05 & 14.59 & 11.69 & 1.84 & 0.17 & 16.44 & 0.39 & 1.08 & 51.8 & Fe CCSN & 1.75 & 2.16 & 1241.87 & 0.18 & 14.43 & 1.34 & 1.01 & 5.74 & 3.91 \\  
35.0 & 0.1 & 16.55 & 12.02 & 2.03 & 0.17 & 16.55 & 0.6 & 1.15 & 51.85 & Fe CCSN & 16.55 &-&-&-&- &-&-& 5.75 & 3.93 \\  
35.0 & 0.15 & 15.76 & 12.24 & 1.84 & 0.19 & 16.92 & 0.36 & 1.04 & 51.79 & Fe CCSN & 1.7 & 2.06 & 1237.36 & 0.18 & 14.99 & 1.33 & 1.05 & 5.8 & 3.92 \\  
35.0 & 0.2 & 17.05 & 13.45 & 2.1 & 0.18 & 17.05 & 0.68 & 1.18 & 51.93 & Fe CCSN & 17.05 &-&-&-&- &-&-& 5.81 & 4.2 \\  
35.0 & 0.25 & 17.46 & 14.06 & 2.14 & 0.18 & 17.46 & 0.69 & 1.19 & 51.95 & Fe CCSN & 17.46 &-&-&-&- &-&-& 5.84 & 4.35 \\  
35.0 & 0.3 & 17.78 & 14.67 & 2.24 & 0.18 & 17.78 & 0.7 & 1.18 & 51.98 & Fe CCSN & 17.78 &-&-&-&- &-&-& 5.84 & 4.49 \\  
35.0 & 0.35 & 18.13 & 15.28 & 2.27 & 0.18 & 18.13 & 0.72 & 1.2 & 52.0 & Fe CCSN & 18.13 &-&-&-&- &-&-& 5.86 & 4.57 \\  
35.0 & 0.4 & 18.48 & 15.86 & 2.31 & 0.18 & 18.48 & 0.74 & 1.22 & 52.04 & Fe CCSN & 18.48 &-&-&-&- &-&-& 5.89 & 4.9 \\  
35.0 & 0.45 & 18.84 & 16.43 & 2.17 & 0.18 & 18.84 & 0.72 & 1.23 & 52.02 & Fe CCSN & 18.84 &-&-&-&- &-&-& 5.9 & 5.25 \\  
35.0 & 0.5 & 18.87 & 16.98 & 2.3 & 0.18 & 18.87 & 0.73 & 1.22 & 52.07 & Fe CCSN & 18.87 &-&-&-&- &-&-& 5.9 & 5.38 \\

\midrule
40.0 & 0.05 & 19.21 & 14.03 & 2.33 & 0.16 & 19.21 & 0.74 & 1.23 & 51.99 & Fe CCSN & 19.21 &-&-&-&- &-&-& 5.84 & 3.93 \\  
40.0 & 0.1 & 19.3 & 14.85 & 2.16 & 0.16 & 19.3 & 0.72 & 1.24 & 52.0 & Fe CCSN & 19.3 &-&-&-&- &-&-& 5.89 & 4.13 \\  
40.0 & 0.15 & 19.71 & 15.52 & 2.34 & 0.16 & 19.71 & 0.77 & 1.26 & 52.05 & Fe CCSN & 19.71 &-&-&-&- &-&-& 5.9 & 4.2 \\  
40.0 & 0.2 & 20.19 & 16.18 & 2.4 & 0.17 & 20.19 & 0.77 & 1.26 & 52.09 & Fe CCSN & 20.19 &-&-&-&- &-&-& 5.92 & 4.31 \\  
40.0 & 0.25 & 20.52 & 16.87 & 2.22 & 0.17 & 20.52 & 0.76 & 1.27 & 52.08 & Fe CCSN & 20.52 &-&-&-&- &-&-& 5.92 & 4.47 \\  
40.0 & 0.3 & 20.8 & 17.56 & 2.35 & 0.17 & 20.8 & 0.72 & 1.25 & 52.15 & Fe CCSN & 20.8 &-&-&-&- &-&-& 5.92 & 4.55 \\  
40.0 & 0.35 & 21.07 & 18.22 & 2.67 & 0.17 & 21.07 & 0.81 & 1.32 & 52.14 & Fe CCSN & 21.07 &-&-&-&- &-&-& 5.96 & 4.88 \\  
40.0 & 0.4 & 21.18 & 18.85 & 2.33 & 0.17 & 21.18 & 0.59 & 1.16 & 52.14 & Fe CCSN & 21.18 &-&-&-&- &-&-& 5.97 & 5.38 \\  
40.0 & 0.45 & 21.18 & 19.45 & 3.66 & 0.17 & 21.18 & 0.59 & 1.16 & 52.09 & Fe CCSN & 21.18 &-&-&-&- &-&-& 5.97 & 5.4 \\  
40.0 & 0.5 & 20.92 & 20.03 & 2.32 & 0.17 & 20.92 & 0.68 & 1.23 & 52.17 & Fe CCSN & 20.92 &-&-&-&- &-&-& 5.97 & 5.42 \\  

\midrule
50.0 & 0.05 & 24.26 & 18.72 & 1.97 & 0.14 & 24.26 & 0.43 & 1.08 & 52.13 & Fe CCSN & 24.26 &-&-&-&- &-&-& 6.02 & 4.03 \\  
50.0 & 0.1 & 24.66 & 19.99 & 2.04 & 0.14 & 24.66 & 0.47 & 1.12 & 52.18 & Fe CCSN & 24.66 &-&-&-&- &-&-& 6.03 & 4.45 \\  
50.0 & 0.15 & 25.46 & 20.76 & 2.1 & 0.14 & 25.46 & 0.45 & 1.1 & 52.2 & Fe CCSN & 25.46 &-&-&-&- &-&-& 6.04 & 4.48 \\  
50.0 & 0.2 & 25.72 & 21.61 & 2.16 & 0.15 & 25.72 & 0.59 & 1.2 & 52.23 & Fe CCSN & 25.72 &-&-&-&- &-&-& 6.02 & 4.54 \\  
50.0 & 0.25 & 25.94 & 22.46 & 2.12 & 0.14 & 25.94 & 0.47 & 1.11 & 52.25 & Fe CCSN & 25.94 &-&-&-&- &-&-& 6.08 & 4.77 \\  
50.0 & 0.3 & 26.14 & 23.22 & 2.16 & 0.15 & 26.14 & 0.63 & 1.21 & 52.28 & Fe CCSN & 26.14 &-&-&-&- &-&-& 6.08 & 5.32 \\  
50.0 & 0.35 & 25.99 & 23.98 & 2.25 & 0.15 & 25.99 & 0.54 & 1.15 & 52.27 & Fe CCSN & 25.99 &-&-&-&- &-&-& 6.08 & 5.41 \\  
50.0 & 0.4 & 25.96 & 24.73 & 2.24 & 0.15 & 25.96 & 0.61 & 1.19 & 52.29 & Fe CCSN & 25.96 &-&-&-&- &-&-& 6.09 & 5.42 \\  
50.0 & 0.45 & 26.1 & 25.47 & 2.14 & 0.15 & 26.1 & 0.54 & 1.17 & 52.3 & Fe CCSN & 26.1 &-&-&-&- &-&-& 6.1 & 5.44 \\  
50.0 & 0.5 & 26.81 & 26.25 & 2.78 &- & 26.81 & 0.43 &-&-& Fe CCSN &-&-&-&-&-&-&-& 6.19 & 5.36 \\ 

\midrule
70.0 & 0.05 & 35.89 & 28.83 & 2.36 & 0.12 & 35.89 & 0.68 & 1.29 & 52.56 & Fe CCSN & 35.89 &-&-&-&- &-&-& 6.26 & 4.49 \\  
70.0 & 0.1 & 36.5 &-& 2.94 &- & 36.5 & 0.58 &-&-& Fe CCSN &-&-&-&-&-&-&-& 7.78 & 4.01 \\ 
70.0 & 0.15 & 37.87 &-& 2.77 &- & 37.87 & 0.55 &-&-& Fe CCSN &-&-&-&-&-&-&-& 8.58 & 4.06 \\ 
70.0 & 0.2 & 41.27 &-& 2.71 &- & 41.27 & 0.52 &-&-& Fe CCSN &-&-&-&-&-&-&-& 8.46 & 3.85 \\ 
70.0 & 0.25 & 43.27 & 33.49 & 2.67 & 0.12 & 43.27 & 0.84 & 1.38 & 52.5 & Fe CCSN & 43.27 &-&-&-&- &-&-& 6.48 & 4.1 \\  
70.0 & 0.3 & 33.86 &-&-&- & 33.86 & 0.1 &-&-& Fe CCSN &-&-&-&-&-&-&-& 17.63 & 8.35 \\ 
70.0 & 0.35 & 36.63 &-& 2.35 &- & 36.63 & 0.36 &-&-& Fe CCSN &-&-&-&-&-&-&-& 9.79 & 4.95 \\ 
70.0 & 0.4 & 36.23 & 36.23 &-&- & 36.23 & 0.29 &-&-& Fe CCSN &-&-&-&-&-&-&-& 9.15 & 5.98 \\ 
70.0 & 0.45 & 36.33 & 36.33 &-&- & 36.33 & 0.23 &-&-& Fe CCSN &-&-&-&-&-&-&-& 11.27 & 5.96 \\ 
70.0 & 0.5 & 36.1 & 36.1 &-&- & 36.1 & 0.34 &-&-& Fe CCSN &-&-&-&-&-&-&-& 11.69 & 5.98 \\ 
\midrule